\documentclass[aps,prb,floatfix,twocolumn,showpacs,preprintnumbers,longbibliography, superscriptaddress]{revtex4-2} 

\usepackage[utf8]{inputenc} 
\usepackage{bibunits}
\usepackage{amssymb,amsfonts,amsmath,mathrsfs,bbold}

\usepackage[normalem]{ulem}

\usepackage[usenames,dvipsnames]{xcolor}
\usepackage[colorlinks=true,citecolor=Blue,urlcolor=Blue,linkcolor=Blue]{hyperref}

\usepackage[all]{hypcap}
\usepackage{multirow}
\usepackage{graphicx}
\usepackage{physics}
\usepackage{dsfont}
\usepackage{braket}
\usepackage{rotating}
\usepackage{soul}

\begin{document}

\title{Two transitions in complex eigenvalue statistics:\\ Hermiticity and integrability breaking
}

\author{Gernot Akemann}
\email{akemann@physik.uni-bielefeld.de}
\affiliation{Faculty of Physics, Bielefeld University, Postfach 100131, 33501 Bielefeld, Germany}
\affiliation{School of Mathematics, University of Bristol, Fry Building, Woodland Road,
Bristol BS8 1UG, UK}

\author{Federico Balducci}
\affiliation{Department of Physics and Materials Science, University of Luxembourg, L-1511 Luxembourg}
\affiliation{Max Planck Institute for the Physics of Complex Systems, N\"othnitzer Str. 38, 01187 Dresden, Germany}

\author{Aur\'elia Chenu}
\email{aurelia.chenu@uni.lu}
\affiliation{Department of Physics and Materials Science, University of Luxembourg, L-1511 Luxembourg}

\author{Patricia P{\"a}{\ss}ler}
\affiliation{Faculty of Physics, Bielefeld University, Postfach 100131, 33501 Bielefeld, Germany}

\author{Federico Roccati}
\affiliation{Department of Physics, Columbia University, New York, New York 10027, USA}
\affiliation{Max Planck Institute for the Science of Light, Staudtstra\ss e 2, 91058 Erlangen, Germany}

\author{Ruth Shir}
\email{ruth.shir@uni.lu}
\affiliation{Department of Physics and Materials Science, University of Luxembourg, L-1511 Luxembourg}


\begin{abstract}
Open quantum systems have complex energy eigenvalues which are expected to follow non-Hermitian random matrix statistics when chaotic, or 2-dimensional (2d) Poisson statistics when integrable. 
We investigate the spectral properties of a many-body quantum spin chain,  the Hermitian XXZ Heisenberg model with imaginary disorder. Its rich complex eigenvalue statistics is found to separately break both Hermiticity and integrability at different scales of the disorder strength. With no disorder, the system is integrable and Hermitian, with spectral statistics corresponding to the 1d Poisson point process. At very small disorder, we find a transition from 1d Poisson statistics to an effective $D$-dimensional Poisson point process, showing Hermiticity breaking. 
At intermediate disorder we find integrability breaking, as infered from the statistics matching that of non-Hermitian complex symmetric random matrices in class AI$^\dag$.
For large disorder, as the spins align, we recover the expected integrability (now in the non-Hermitian setup), indicated by 2d Poisson statistics.  
These conclusions are based on fitting the spin chain data of numerically generated nearest and next-to-nearest neighbour spacing distributions to an effective 2d Coulomb gas description at inverse temperature $\beta$. 
We confirm such an effective description of random matrices also applies in class AI$^\dag$ and AII$^\dag$ up to next-to-nearest neighbour spacings.
\end{abstract}

\maketitle


\section{Introduction}

One of the most studied indicators of quantum chaos in isolated systems is the spectral statistics of nearest-neighboring eigenvalues. 
The quantum chaos conjectures by Berry and Tabor (BT)~\cite{berry1977level} and Bohigas, Giannoni and Schmit (BGS)~\cite{bohigas1984characterization}, cf.~\cite{casati1980}, 
identify the nearest-neighbour spectral statistics as a tool to distinguish generic closed quantum chaotic systems from integrable ones.
BT conjectured that generic closed quantum integrable systems have spectral statistics corresponding to the statistics of a Poisson point process in one dimension (1d Poisson). 
BGS conjectured that closed quantum chaotic systems follow spectral statistics of random matrices and are classified according to their symmetry classes. These conjectures were shown to work in many closed quantum systems~\cite{DAlessio2016Quantum} and BGS was derived from a semi-classical expansion \cite{muller2004semiclassical}, cf.~\cite{haake1991quantum}. 
The crossover between chaos and integrability, where an integrability breaking term is added to an otherwise integrable model, or an integrability restoring term is added to a chaotic model, were also studied~\cite{DAlessio2016Quantum, PhysRevLett.122.180601}.

Quantum systems are in general not perfectly isolated, but rather interact with their surrounding environment. 
When the timescales of the environment are much shorter than the system's ones and under the Born approximation, 
the reduced system's dynamics can be effectively described by a Schr\"odinger equation with a non-Hermitian Hamiltonian~\cite{MingantiPRA2019, NaghilooNatPhys2019,RoccatiOSID2022} and thus generically has a complex spectrum. 
 Grobe, Haake and Sommers (GHS)~\cite{grobe1988quantum} conjectured that, for Markovian, open quantum systems, spectral properties can be used to distinguish between chaos and integrability, in a way similar to the Hermitian case. In particular, open quantum systems that are chaotic should display non-Hermitian random matrix statistics, while integrable systems with complex spectra should follow the statistics of a Poisson point process in two dimensions (2d Poisson). 
This conjecture was tested early on for the discrete quantum map of periodically kicked tops with damping \cite{grobe1988quantum}, and in random neural networks \cite{sompolinsky1988chaos}, see below for further recent examples. In both cases, the nearest-neighbour 
spacing distribution was found to agree with 2d Poisson in the integrable case, and with that of the complex Ginibre ensemble \cite{ginibre1965} in the fully chaotic regime. 

Recently, the question of which symmetry class of random matrices is appropriate for dissipative systems has been reopened and explored in greater detail \cite{hamazaki2020universality}, revisiting the previously suggested Bernard-LeClair classification \cite{bernard2002classification,magnea2008random}. Based on numerical simulations, it was suggested that only three different symmetry classes of local bulk statistics exist \cite{hamazaki2020universality}, with representatives being class A (complex Ginibre) \cite{ginibre1965}, AI$^\dag$ (complex symmetric) and AII$^\dag$ (complex self-dual). 
Several authors have provided further examples for the non-Hermitian quantum chaos conjecture of GHS, finding a crossover from 2d Poisson to one of these random matrix statistics in quantum systems. These include examples from 
2d random Schr\"odinger operators \cite{hatano1996localization}, 4d lattice quantum field theory with chemical potential \cite{markum1999non,kanazawa2021new},  directed complex networks \cite{ye2015spectral}, 
quantum spin chains \cite{akemann2019universal,hamazaki2020universality,Rubio2022chaos}, kicked rotors \cite{jaiswal2019universality}, 2d Anderson models with disorder \cite{huang2020anderson}, 
and beyond physics in theoretical ecology \cite{Akemann_2024}.

The crossover between integrable and chaotic systems has thus currently been studied in systems that are either Hermitian---it then corresponds to going from a 1d Poisson process to a Hermitian random matrix ensemble---or non-Hermitian---the transition is then from a 2d Poisson to one of the non-Hermitian symmetry classes. However, the interplay between Hermiticity breaking and integrability breaking has never been investigated. We do it here, by studying 
a strongly-interacting many-body system with dissipative local disorder. In absence of dissipation, disordered interacting quantum models have been the focus of much attention recently, due to the presence of many-body localized phases~\cite{Basko2006Metal,Oganesyan2007Localization,Znidaric2008Many,Pal2010Many,Bardarson2012Unbounded,DeLuca2013Ergodicity,Abanin2019Colloquium,Sierant2024Many}. These are robust, non-ergodic phases of quantum matter, that violate the eigenstate thermalization hypothesis~\cite{Deutsch1991Quantum,Srednicki1994Chaos} and display an emergent integrability via local integrals of motion~\cite{Serbyn2013Local,Ros2015Integrals,Imbrie2016Diagonalization}. Similar non-chaotic regimes have been conjectured to be present also in an open setting, especially in the context of non-Hermitian Hamiltonians~\cite{hamazaki2019non,Zhai2020Many,Suthar2022NonHermitian,Ghosh2022Spectral,detomasi2023stable}.

The model considered in this work, introduced in Ref.~\cite{roccati2024diagnosing}, has the peculiarity to have a purely imaginary disorder, breaking at the same time both integrability and Hermiticity. This model is motivated by a recent experimental work~\cite{lapp2019engineering}, showing that local tunable loss terms can be engineered in synthetic lattices. Moreover, the model allows us to investigate whether the breaking of Hermiticity happens on a different scale than the transition between chaotic and integrable behaviour. We quantitatively characterize the crossovers by comparing the 
nearest-neighbour (NN) level statistics of complex eigenvalues to that of a static Coulomb gas in 2d. In order to go beyond the smallest local scale, we will also use next-to-nearest neighbour (NNN) spacing distributions. The 2d Coulomb gas (2dCG) allows us to reproduce the 2d Poisson, and approximate the  random matrix class AI$^\dag$ as a function of the Coulomb gas inverse temperature $\beta$ \cite{akemann2022spacing}, thus enabling to capture a transition from integrable ($\beta=0)$ to chaotic ($\beta=1.4$) behavior, respectively. The breaking of Hermiticity is captured through a $D$-dimensional Poisson point process, with $D$ transitioning from a value of 1 for a purely Hermitian spectrum to 2 for strongly non-Hermitian systems.

The article is organised as follows. The next Section \ref{Sec:model} describes our model, the XXZ spin chain with purely imaginary disorder, which is Hermitian and integrable at vanishing disorder, and its symmetries. 
The main Section \ref{Sec:Results} contains our results, namely the transition between random matrix and 2d Poisson statistics in $\beta$ of the 2dCG, and, for very small disorder, the transition from 1d to 2d Poisson statistics, indicating a Hermiticity breaking crossover. In Section \ref{Sec:2d_Spectral}, we introduce the corresponding complex eigenvalue statistics for NN and NNN 
spacing distributions 
used to characterize the model, namely first, an interpolating 2dCG ensemble for which an accurate surmise exists close to 2d Poisson, 
second, Poisson point process statistics (in varying dimension $D$), and third, the random matrix statistics of class AI$^\dag$, appropriate for the considered model. 
Section \ref{Sec:methods} describes our methods, in particular the unfolding procedure used in 2d and the tools to measure distances between two probability distributions.  For completeness, we briefly present complex spacing ratios introduced by \cite{sa2020complex} as an alternative tool. 
 Our conclusions and open questions are presented in Section \ref{Sec:concs}.
Appendix \ref{App:RMT} summarises the other two distinct random matrix symmetry classes A and AII$^\dag$, including NN and NNN spacings.

\section{XXZ model with dissipative disorder}\label{Sec:model}
This section describes the dissipative model we consider, together with some aspects of its symmetries and spectrum.

\begin{figure}
    \centering
    \includegraphics[scale=0.5]{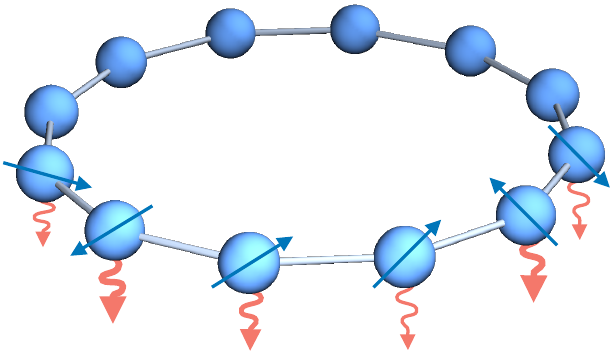}
    \caption{Schematic illustration of the spin chain model with local dissipative disorder described by Eq. \eqref{XXZ_iW_Hamiltonian}.}
    \label{fig:XXZiW}
\end{figure}

\subsection{Definition of the model}

We consider a chain of $L$ interacting spins with local dissipative disorder, described by the non-Hermitian Hamiltonian
\begin{equation}
    \label{XXZ_iW_Hamiltonian}
    H = H_{\textrm{XXZ}} - i \, \Gamma/2,
\end{equation}
with the XXZ spin chain Hamiltonian 
\begin{equation}\label{eq:H_XXZ}
    H_{\textrm{XXZ}} = \sum_{a=1}^L \left( S_a^x S_{a+1}^x + S_a^y S_{a+1}^y + \Delta S_a^z S_{a+1}^z \right),
\end{equation}
and 
\begin{equation}
    \label{Gamma}
    \Gamma = \sum_{a=1}^L \gamma_a \left(S^z_a + \frac{1}{2} \mathds{1}\right) 
\end{equation}
representing site-dependent, local, \textit{dissipative} disorder, because of the imaginary $i$ in \eqref{XXZ_iW_Hamiltonian}. 
The site-dependent losses $\gamma_a$ can be experimentally implemented, as demonstrated in a synthetic lattice \cite{lapp2019engineering}. Here, they are sampled from a uniform distribution $\mathcal{U}(0,\gamma)$, where $\gamma$ represents the `strength' of the disorder. For each value of $\gamma$, we generate spectral data from a number of realizations of the model, each realization having a different randomly chosen set of local disorder terms, $\{\gamma_a \}_{a=1}^L$.  
We consider the case of periodic boundary conditions, as depicted in Figure~\ref{fig:XXZiW}. 

Note that in this model, all the disorder and its implications to the spectral statistics come from the openness of the system---the spin chain Hamiltonian \eqref{eq:H_XXZ} is always fixed and is integrable according to the Bethe Ansatz \cite{XXZ_bethe_ansatz}. For the study of spectral statistics of non-Hermitian models with real disorder see e.g.~\cite{hamazaki2019non, huang2020anderson, roccati2024diagnosing}.
In~\cite{roccati2024diagnosing}, it was shown that increasing the range $\gamma$, the statistics of the \textit{singular values} exhibit a crossover from chaotic to integrable.  
A similar observation is made here from the eigenvalue spacing statistics. We complete the story by extending the analysis to very small disorder, which reveals a richer behavior coming from Hermiticity breaking. 
For some aspects of thermalization and transport in this model, see \cite{Mahoney:2024sgu}. 
This model preserves the total spin  in the $z$ direction, $\mathcal{S}_z = \sum_{a=1}^L S_a^z$. We thus work in a specific total $\mathcal{S}_z$ magnetization sector, usually the largest one with $\mathcal{S}_z=0$ which occurs for even $L$ and is of dimension $\binom{L}{L/2}$. Without disorder ($\gamma=0$), additional symmetries need to be taken into account to avoid exact degeneracies in the spectrum (see e.g.~\cite{joel2013introduction}). Specifically, each $\mathcal{S}_z$ sector then 
splits into two: one sector with states of even Parity and another with states of odd Parity.

The Hamiltonian~\eqref{XXZ_iW_Hamiltonian} falls into the universality class AI$^\dagger$ of non-Hermitian random matrices~\cite{hamazaki2020universality}. These are complex symmetric matrices which satisfy $H=H^T$, with $H \neq H^\dagger$.
In addition, the non-Hermitian Hamiltonian~\eqref{XXZ_iW_Hamiltonian} is pseudo-Hermitian, as shown in Section~\ref{Sec:PH_spectrum}, and thus its complex eigenvalue spectrum is symmetric across a line parallel to the real axis of the complex plane at $(-i/4)\sum_{a=1}^L\gamma_a$.


\subsection{Pseudo-Hermitian complex spectrum}
\label{Sec:PH_spectrum}

The complex spectrum of the Hamiltonian~\eqref{XXZ_iW_Hamiltonian} has reflection symmetry along a line parallel to the real axis, i.e. the eigenvalues come in complex conjugate pairs above and below this line. This reflection symmetry in the complex spectrum is a result of the matrix being \textit{pseudo-Hermitian}~\cite{pauli1943dirac, feinberg2022metrics, melkani2023degeneracies}.  A matrix is pseudo-Hermitian if there exists an invertible matrix $S$ such that
\begin{equation}
    \label{Pseudo-Hermitian}
    H = SH^\dag S^{-1} ~.
\end{equation}
Indeed, if such a similarity transformation exists, 
 and $\lambda$ is an eigenvalue of $H$ with right eigenvector $|R\rangle$, $H|R\rangle = \lambda |R\rangle$, then also $\lambda^*$ is an eigenvalue of $H$. This can be shown by taking the Hermitian conjugate $H^\dag|R^*\rangle = \lambda^* |R^*\rangle$ and using~\eqref{Pseudo-Hermitian}, giving $HS|R^*\rangle = \lambda^* S|R^*\rangle$, which means that $\lambda^*$ is an eigenvalue of $H$ with eigenvector $S|R^*\rangle$.

In our case, such an $S$ would need to flip the sign of $\Gamma$ in Eq.~\eqref{Gamma} \footnote{After removing the term proportional to the identity, which is responsible for the constant imaginary shift in the complex plane.}, while keeping $H_{\textrm{XXZ}}$ unchanged.  It turns out that this can be achieved by the operator
\begin{equation}
    S = \prod_{a=1}^L S_a^x ~,
\end{equation}
which acts as $S S^x_a S^{-1}=+S^x_a$, $S S^y_a S^{-1}=-S^y_a$ and $S S^z_a S^{-1}=-S^z_a$ for any site $a$. In $H_{\textrm{XXZ}}$, the spin operators appear in pairs and so the sign change is cancelled. In turn,  the $S^z_a$ operators appear once in each term  of $\Gamma$, leading to $S \Gamma S^{-1} = -\Gamma$, as needed to fulfill the symmetry \eqref{Pseudo-Hermitian}.


\section{Results for the spin chain}
\label{Sec:Results}
We now characterize the spectral statistics of the spin chain with dissipative disorder described by the non-Hermitian Hamiltonian~\eqref{XXZ_iW_Hamiltonian}.  
We do so by comparing the distributions of its eigenvalue spacings with that of the 2d spectral statistics of the reference models, detailed in Sec.~\ref{Sec:2d_Spectral}, using the methods presented in Sec.~\ref{Sec:methods}.
We focus on the NN and NNN spacing distributions for values of the disorder strength from $\gamma=0-20$, fitting the numerical distribution to either the 2dCG distributions and finding the corresponding $\beta$, or to a $D$-dimensional Poisson point process with $1\leq D \leq 2$. For $L=16$, we present NN, NNN, and complex spacing ratio data for over 450 disorder realizations, while for $L=12,14$, we show this data for 1000 disorder realizations. To complete the picture, a qualitative discussion of the complex spacing ratios \cite{sa2020complex} is included at the end of the section. 

We expect the spectral statistics of our model to behave similarly to those of a XXZ spin-chain Hamiltonian with real disorder, which exhibits a crossover of the spectral statistics from chaos to integrability (for non-zero disorder strength) as the disorder strength is increased. 
In such a case, the NN spectral statistics changes from random matrix statistics approximated by the Wigner surmise $\propto s e^{-a s^2}$ for the GOE in this model to 1d Poisson, $\propto e^{-s}$, $s$ being the spacing between NN eigenergies. 
In 1d, there is universal level repulsion between the eigenvalues only in the first, chaotic regime, while in the second, 1d Poisson displays uncorrelated eigenvalues---a universal signature of generic integrable (deterministic) systems.  

\begin{figure*}
    \includegraphics[width=\linewidth]{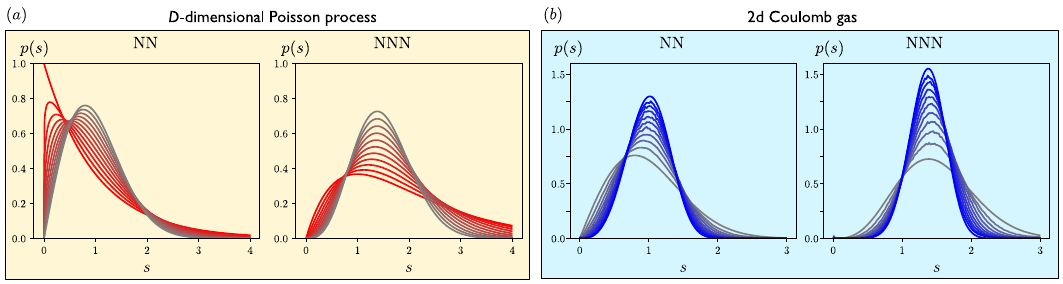}
    \caption{\textbf{(a)} Spacing distributions of the Poisson process,  with dimension $1\leq D\leq 2$ varying in steps $0.1$ (red--grey). Notice that in both cases the maximum moves from right to left when decreasing dimension $D$. \textbf{(b)} The numerically generated NN and NNN spacing distributions of the 2dCG  for $\beta=0-2$, increasing in steps $0.2$ (grey--blue). This also includes analytic formulas for 2d Poisson with $\beta=0$, \eqref{PoiD-NN} and \eqref{PoiD-NNN} with the lowest maxima, and random matrix class A with $\beta=2$, \eqref{ANN} and \eqref{ANNN}, with highest maxima. 
    Note that the grey curves for $D=2$ in (a) are the same as the grey curves for $\beta=0$ in (b) since the two processes then coincide.  
    \label{Fig:2d-statistics}}
\end{figure*}

In the non-Hermitian (2d) case with imaginary disorder, where the eigenvalues lie in the complex plane, we expect the spectral statistics to exhibit a crossover between random matrix statistics of class AI$^\dag$ in our case to that of the 2d Poisson point process---which has a linear repulsion from the area measure, see \eqref{PoiD-NN} at $D=2$. 
We follow the proposal  \cite{akemann2019universal} and use the spectral statistics of a 2d Coulomb gas with some value for $\beta\geq0$, which includes 2d Poisson at $\beta=0$ as a special case and approximates the AI$^\dag$ class for $\beta \approx 1.3-1.4$ \cite{akemann2022spacing}. Fitting $\beta$ thus allows us to capture this crossover. 
However, the transition has more structure than initially expected, showing two different crossovers, as we summarise below and detail in subsequent subsections.

\begin{figure}[b!]
    \centering
    \includegraphics[width=\linewidth]{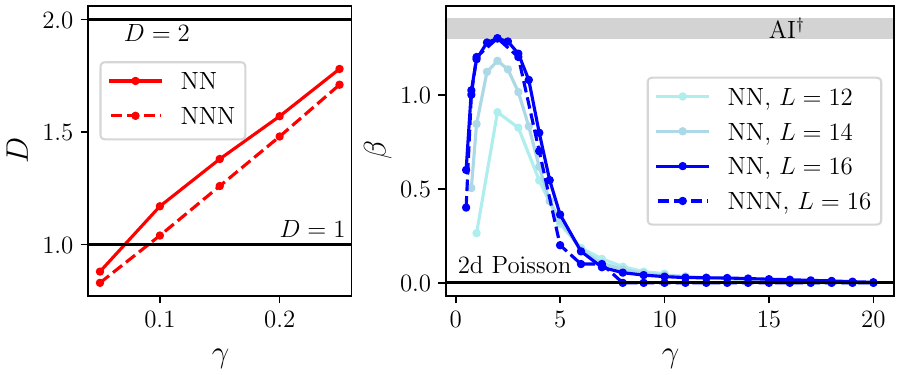}
    \caption{Two transitions in the XXZ model with dissipative disorder. \textbf{Hermiticity breaking (left):} For $0.05\leq\gamma<0.25$ we find that the NN and NNN spectral statistics changes from those of 1d Poisson to 2d Poisson, thus Hermiticity is broken first. Shown is the fitted effective dimension $D$ for NN (full curve) and NNN (dashed curve). Data is shown for $L=16$ spins in the $\mathcal{S}_z = 0$ magnetization sector.
    \newline
    \textbf{Integrability breaking (right):}
    For larger $\gamma$, i.e. $0.5<\gamma<20$ we find a very good fit to $\beta$ of the 2dCG. 
    At $\gamma=2$ the NN (full curve) and NNN statistics (dashed curve) are very close to those of AI$^\dagger$ corresponding to $\beta\approx 1.3-1.4$ (horizontal band),  
    representing a complete breaking of integrability. As $\gamma$ is increased further, there is a crossover to the statistics of 2d Poisson (horizontal line at $\beta=0$).}
    \label{fig:summary}
\end{figure}
\begin{figure*}[t]
    \centering
	\includegraphics[width=\linewidth]{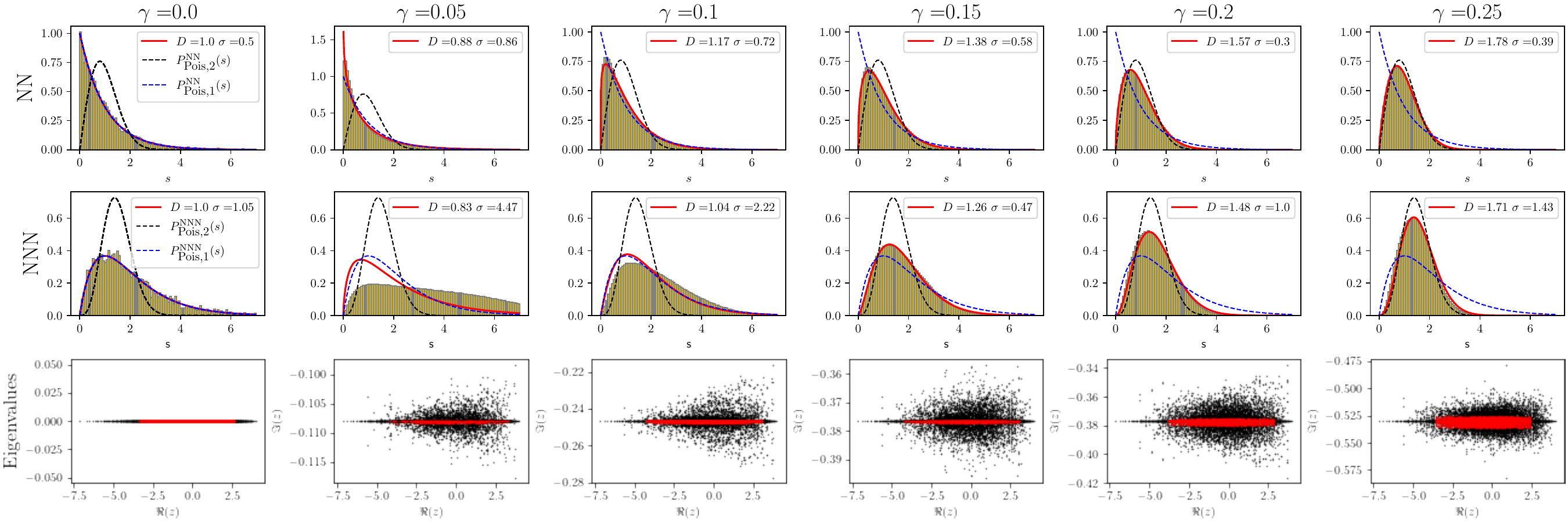}
   \caption{The NN (top), NNN spacing distribution (middle) of at least 490 disorder realizations, and scatter plot with selected eigenvalues highlighted in red for a single realization of the spectrum (bottom) for $\gamma=0,0.05,0.1,0.15,0.2,0.25$. In red (full line) we show \eqref{PoiD-NN} and \eqref{PoiD-NNN} for the NN and NNN distribution, respectively, for the best fitted value for the effective dimension $D$, with corresponding standard deviation $\sigma$ (in units of $10^{-2})$.  For comparison we also show the results for $D=1$ (blue dashed) and for $D=2$ (black dashed line) as an orientation. For pure XXZ, $\gamma=0.0$, the spectrum is real, and since there is no disorder the data shown is for a single realization. In addition, the full $\mathcal{S}_z = 0$ is degenerate and splits into two sectors: an even Parity sector and an odd Parity one; we show statistics only for the even Parity sector.  
    Switching on $\gamma>0$ we see how these sectors start to mix, which may also be responsible for the disagreement of the NNN at the smallest values of $\gamma=0.05$.
    }
    \label{Fig:smallga}
\end{figure*}

On a technical level, the spin chain data need to be drawn away from the edges of the eigenvalue distributions, where the statistics may be different or not even universal, as discussed at the beginning of Section \ref{Sec:2d_Spectral}. To do so, we focus on the bulk given by the denser inner part of the spectrum. We systematically select only that part of the spectrum for which the eigenvalue density is not less than some fraction of the maximal density. For $L=16$ spins with $\gamma=0.75 - 20$, we use $1/2$ of the maximal density;  for $L=14$ and $L=12$, we use a $1/3$ of the maximal density as a cutoff for the eigenvalues to be included in the analysis. For $L=16$ spins with $\gamma < 0.75$ we choose a cutoff equal or smaller to $1/4$ the maximal density because, as we discuss below, the spectrum is very peaked around a line parallel to the real line.\\

\noindent

The 2d spectral statistics of the $D$-dimensional process and 2dCG,  detailed in Section \ref{Sec:2d_Spectral}, are presented in Fig.~\ref{Fig:2d-statistics}. For our analysis, the relevant features are as follows. 
The NN distribution of the 2dCG has its leftmost and lowest maximum at $\beta=0$, see Fig.~\ref{Fig:2d-statistics}(b) left (grey curve). Increasing $\beta$ leads to a monotonous increase of the maximum in height and in position to the right (grey to blue). The random matrix class AI$^\dag$ expected at maximal chaos is the one with the lowest maximum of the three distinct random matrix classes (see Fig.~\ref{Fig:3RMTs}), reached at approximately $\beta\approx 1.3-1.4$. As such, it is approached by the 2dCG monotonously from below, starting at $\beta=0$. 
What happens if the maximum of the NN spacing distribution of our spin chain data---which is normalised and has first moment of unity---is to the \textit{left} of the 2d Poisson distribution (at $\beta=0$)? Obviously the 2dCG  
will no longer provide a good fit then.  Only the $D$-dimensional Poisson distribution might provide a good fit then for $D<2$, see Fig. \ref{Fig:2d-statistics}(a) left.  There, it is shown that the maximum of the distribution goes from left to right as the effective dimension is increased.

Our findings can be summarised by Fig. \ref{fig:summary} and described as follows. In the small disorder regime, there is a crossover between 1d Poisson at $\gamma=0$---where our XXZ spin chain is integrable--- towards 2d Poisson, that is well described by an increase in effective dimension $1\leq D<2$. We reach up to $D\approx 1.7$ at $\gamma=0.25$ for both NN and NNN spacing distributions. 
Therefore, Hermiticity is broken first on a smaller scale.  This is summarised in  Fig.~\ref{fig:summary} left and described in detail in Subsection \ref{Sec:Hbreak}.

For intermediate to large values of $\gamma=0.5 - 20$, we see a rise from $\beta\approx0.6$ at $\gamma=0.5$ to $\beta\approx1.3-1.4$ at $\gamma=2$ where class AI$^\dag$ is located, to which we also compare. We see that integrability is broken as we approach the fully chaotic behaviour characterised by random matrix class AI$^\dag$. For larger $\gamma$, the disorder term \eqref{Gamma} leads to an alignment of spins and thus again to a deterministic behaviour, approaching 2d Poisson with $\beta=0$ around $\gamma\approx 8$, see Fig. \ref{fig:summary} right. 
This is discussed in Subsection \ref{Sec:Ibreak}, including the approach to class AI$^\dag$ with system size $L$ in Subsection \ref{Sec:Ldependence}. 
The asymptotic approach to $\beta=0$ between $\gamma=8-20$ is discussed in Subsection \ref{Sec:AsymptPoisson}, where we also use the surmise \eqref{Ps_2dCG} for the finer scaling in the NN distribution. To complete the picture, we show and discuss results for the complex spacing ratios at various values of $\gamma$ in Subsection \ref{Sec:XXZiW_CSR}.

\subsection{Hermiticity breaking at small disorder $\gamma$: \\from 1d to 2d Poisson}
\label{Sec:Hbreak}

This subsection presents the regime of small $\gamma$, with range $\gamma=0-0.25$, and compares our model to Poisson statistics with varying dimension, Eqs.~\eqref{PoiD-NN} and \eqref{PoiD-NNN} for the NN and NNN spacing distribution, respectively, see Fig.~\ref{Fig:smallga}. The corresponding spectrum is also shown---notice the different scales on the $y$-axis---with the part of the spectrum that is used, representing the bulk, highlighted in red. As explained in Section \ref{Sec:model}, the spectrum is symmetric with respect to a line parallel to the real axis. This symmetry line changes with $\gamma$.

\begin{figure}[b!]
    \centering
    \vspace*{-0.3cm}
    \includegraphics[width=\linewidth]{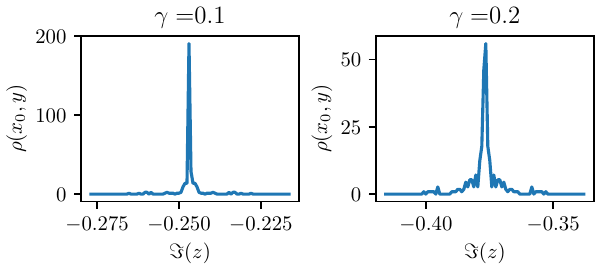}
    \caption{Cuts through the spectral density (for a single realization), in imaginary direction  at $x_0\approx 0$, for $\gamma=0.1$ (left) and $\gamma=0.2$ (right).   \label{Fig:cuts} 
    }
\end{figure}

Starting with $\gamma=0$ (left plots), the spectrum is real, the XXZ model integrable,  
and we obtain a very good agreement with 1d Poisson statistics for both NN and NNN as expected. 
Notice that at 
$\gamma=0$ the energies are doubly degenerate, which is why we only show spacings with $\mathcal{S}_z = 0$ and parity $P=+1$ (for $L=16$, this sector is of dimension 6470). The plots for $\mathcal{S}_z = 0$ with $P=-1$ (for $L=16$, this sector is of dimension 6400) look similar. 
For our smallest value of $\gamma$, we obtain a somewhat reasonable fit for the NN, with $D\approx 0.88$ consistent with $D=1$. In contrast, the distribution for the NNN does not yield a good fit at all, a feature that persists for $\gamma=0.1$. A possible explanation is that for these small values of $\gamma$ the spectrum is almost singular, being concentrated on a line parallel to the real line, with very few eigenvalues in the complex plane, see Fig.~\ref{Fig:cuts}, and the unfolding procedure described in Section \ref{Sec:unfolding} for a 2d complex spectrum may then not be completely suitable. 
For $\gamma=0$, we used the 1d version of the unfolding method described in \ref{Sec:unfolding} \footnote{We found that for 1d, $\Sigma = 1.8 \bar{s}$ gives good results.}.

\begin{figure*}
		\hspace{5pt}{$\gamma=0.5$} \hspace{45pt}{$\gamma=1$}\hspace{47pt}{$\gamma=2$} \hspace{47pt}{$\gamma=3$}\hspace{47pt}{$\gamma=4$}\hspace{47pt}{$\gamma=5$} \hspace{47pt}{$\gamma=10$}\hfill\\
\begin{turn}{90}\hspace{20pt}{NN}\end{turn}
   	\includegraphics[width=0.135\linewidth,angle=0]{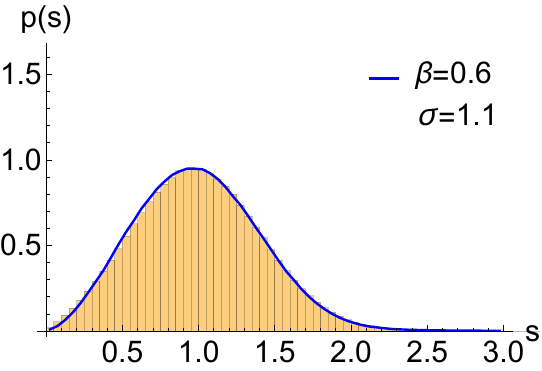}
	\includegraphics[width=0.135\linewidth,angle=0]{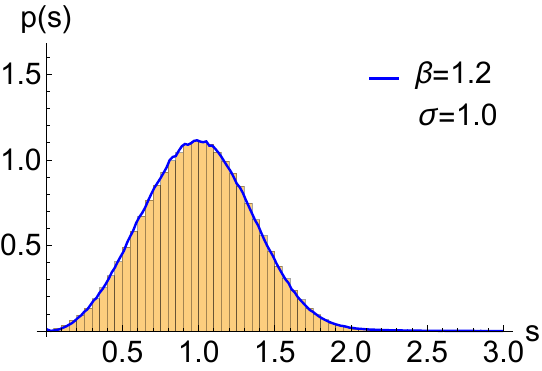}
	\includegraphics[width=0.135\linewidth,angle=0]{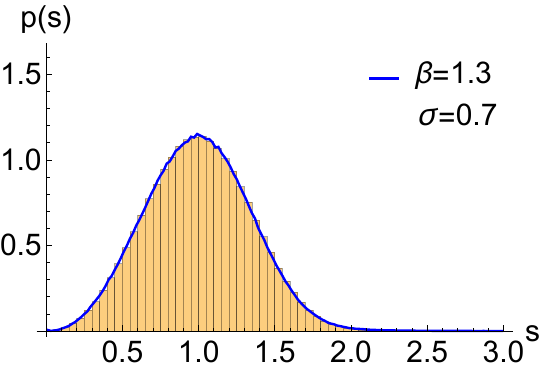}
	\includegraphics[width=0.135\linewidth,angle=0]{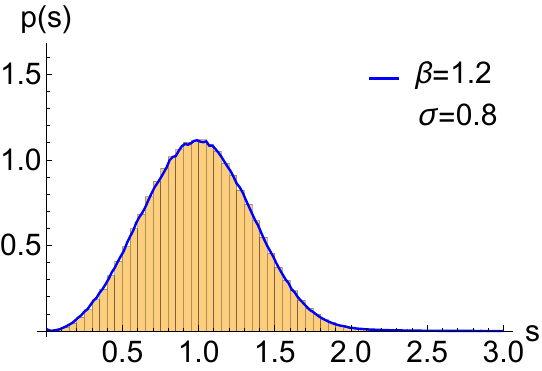}
	\includegraphics[width=0.135\linewidth,angle=0]{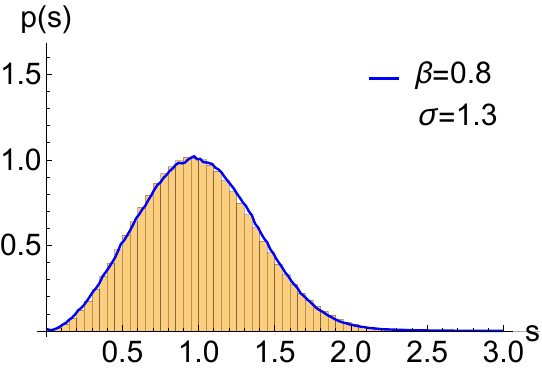}
   	\includegraphics[width=0.135\linewidth,angle=0]{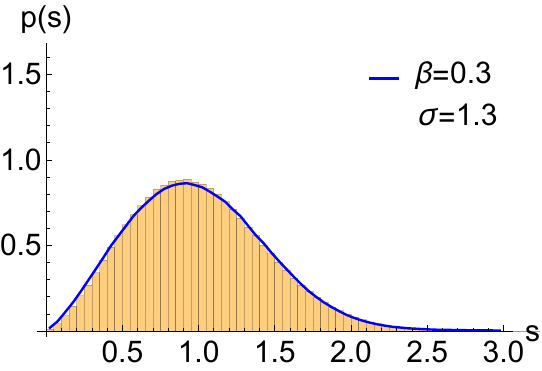}
	\includegraphics[width=0.135\linewidth,angle=0]{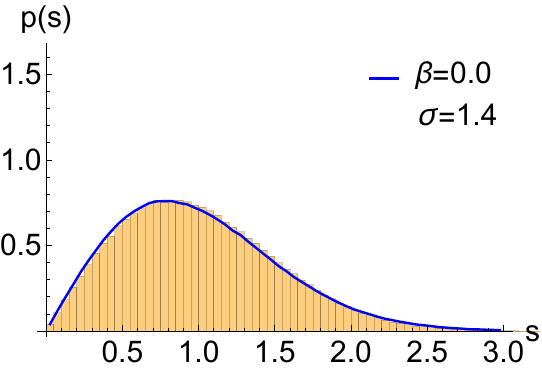}	
\begin{turn}{90}\hspace{15pt}{NNN}\end{turn}
   	\includegraphics[width=0.135\linewidth,angle=0]{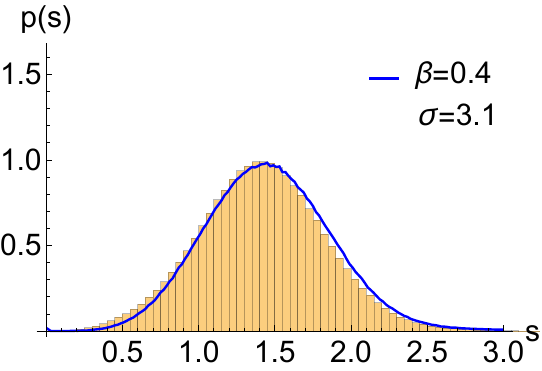}
	\includegraphics[width=0.135\linewidth,angle=0]{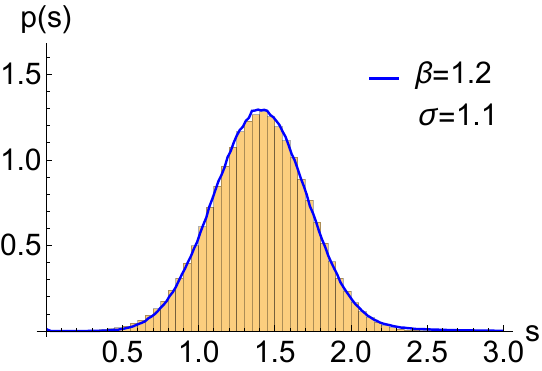}
	\includegraphics[width=0.135\linewidth,angle=0]{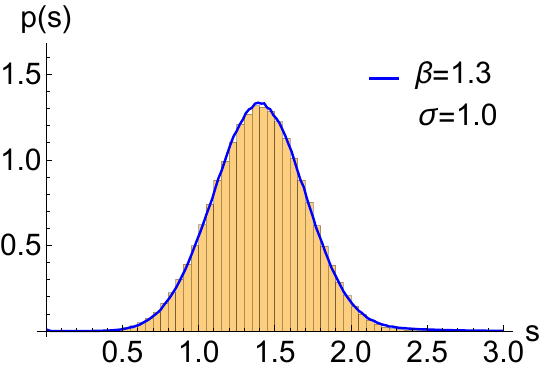}
	\includegraphics[width=0.135\linewidth,angle=0]{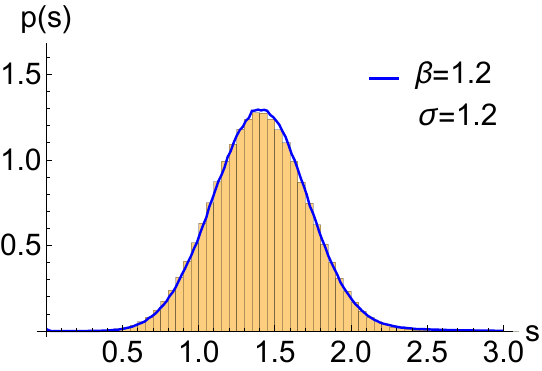}
	\includegraphics[width=0.135\linewidth,angle=0]{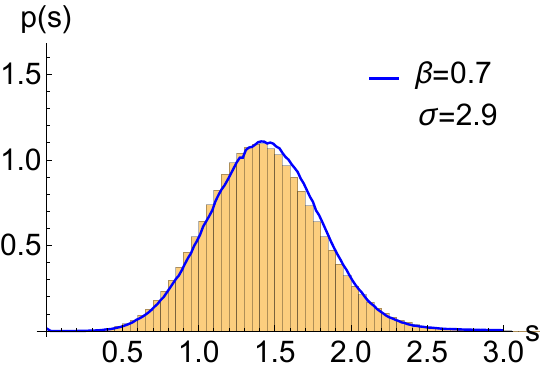}
   	\includegraphics[width=0.135\linewidth,angle=0]{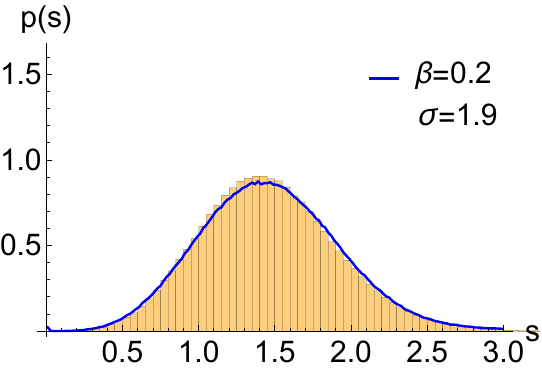}
	\includegraphics[width=0.135\linewidth,angle=0]{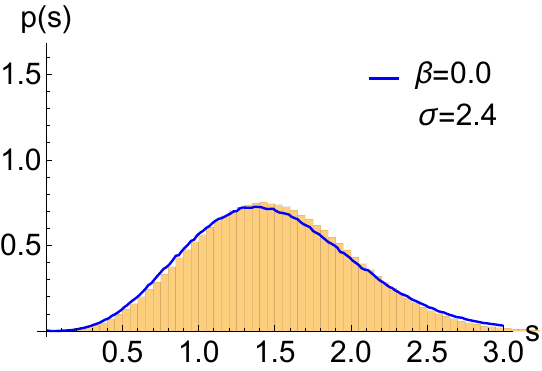}
 
    \begin{turn}{90}\hspace{7pt}{Eigenvalues}\end{turn}
    	\includegraphics[width=0.135\linewidth,angle=0]{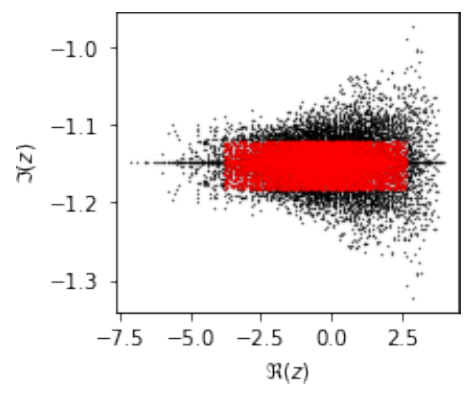}
	\includegraphics[width=0.135\linewidth,angle=0]{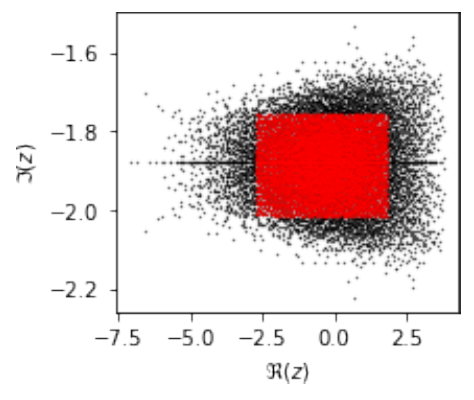}
	\includegraphics[width=0.135\linewidth,angle=0]{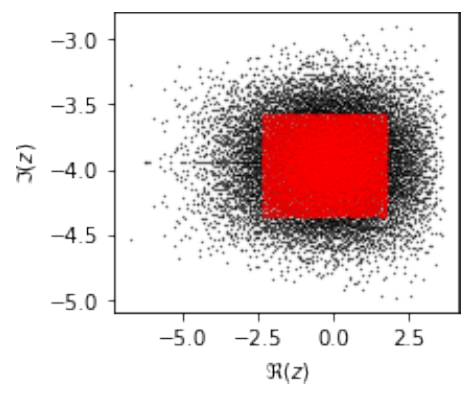}
	\includegraphics[width=0.135\linewidth,angle=0]{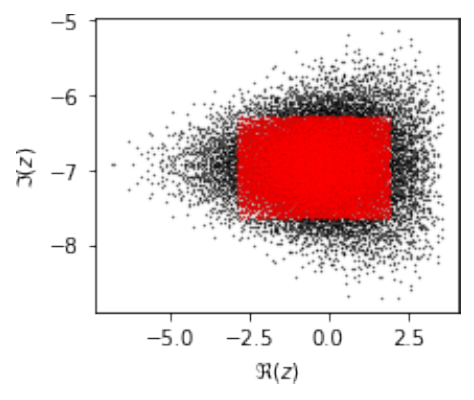}
	\includegraphics[width=0.135\linewidth,angle=0]{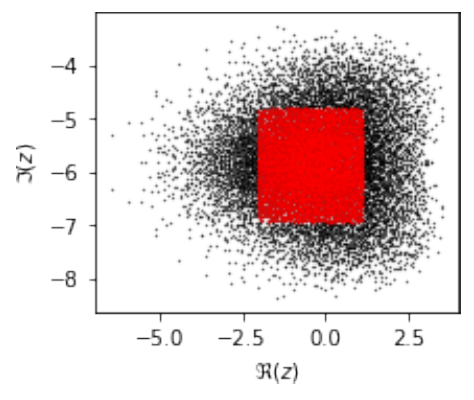}
   	\includegraphics[width=0.135\linewidth,angle=0]{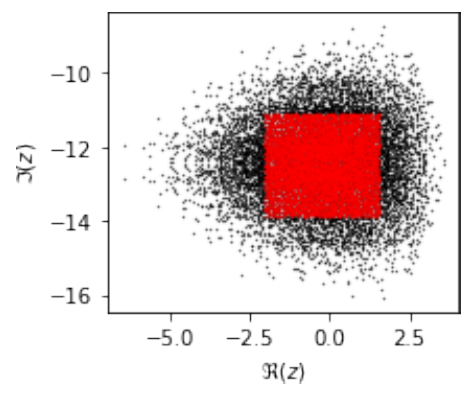}
	\includegraphics[width=0.135\linewidth,angle=0]{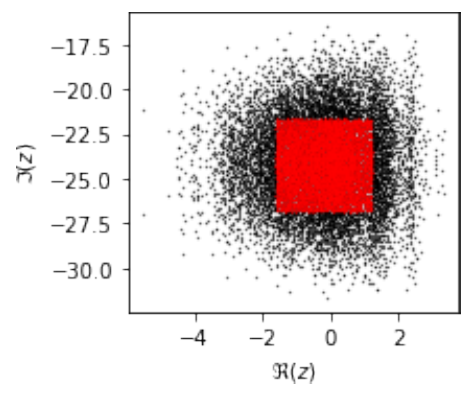}	
	\caption{Comparison of the model \eqref{XXZ_iW_Hamiltonian} with increasing dissipative order $\gamma$, with size $L=16$ and spins in the $\mathcal{S}_z = 0$ magnetization sector (histograms) for at least 450 realizations, and the best fit to the 2dCG (blue curves) of size $N=5000$. The NN distribution is given in the first row, whereas the NNN spacing is shown in the second row. Additionally, we give the standard deviation $\sigma$ in units $10^{-2}$ for each comparison of the spin chain data with the 2dCG. In the bottom row we show an example of the complex eigenvalue data for a single realization, with the points used in the statistics highlighted in red.}
 \label{Fig:intermedg}
\end{figure*}

From $\gamma=0.1$ both NN and NNN start to deviate from $D=1$, and the best fits in $D$ reach values up to $D\approx 1.7$ at $\gamma=0.25$. Comparing  NN and NNN leads to consistent values within a small margin. 
We thus believe that the fits to a one-parameter family of distributions are not a coincidence, but rather capture properties of an underlying effective Poisson interaction.

This leads to the interpretation that Hermiticity is broken first, while maintaining an (almost) integrable system, the characteristic of which is given by a Poisson distribution at intermediate dimension $1<D<2$. Increasing to the next higher value of $\gamma=0.5$, detailed in the next subsection \ref{Sec:Ibreak}, the maximum of the NN crosses to the right of the 2d Poisson  distribution (at $\beta=0$), which then fits the 2dCG. 
The fact that $D=2$ is not fully reached by increasing $\gamma$ here can be interpreted as a competition between Hermiticity and integrability breakings around the scale $\gamma=0.25-0.5$.

As an example for the crossover between one- and two-dimensional spectra, we show in Fig. \ref{Fig:cuts} cuts through the spectral density perpendicular to the real axis at $x_0 \approx 0$ as a function of imaginary part $y$. Noting the difference in scale, we see that between $\gamma=0.1$ and $0.2$ shown here, the spectrum slowly softens and starts spreading into the complex plane. The difference in abundance of eigenvalues with non-vanishing imaginary part may explain why the NNN distribution in Fig. \ref{Fig:smallga} fits well only from a certain value of $\gamma=0.15$ onwards.


\begin{table*}[]
    \centering 
	\caption[]{List of best fit $\beta$ for the comparison of the NN spacing  and NNN spacing 
 of the spin chain with $L=16$ and $\gamma=0.5,0.75,1,2,\ldots 20$ to the 2dCG and the corresponding standard deviation $\sigma$ in units of $10^{-2}$. Notice that apart from $\gamma=0.5$ the values obtained for $\beta$ from NN and NNN agree or differ by $0.1$ only. }\vspace{3mm}
	\begin{tabular}{|c|c|c|c|c|c|c|c|c|c|c|c|c|c|c|c|c|c|c|c|c|c|c|c|}
		\hline
		&$\boldsymbol{\gamma}$& \textbf{0.5} & \textbf{0.75} &\textbf{1.0}  & \textbf{2.0} &\textbf{3.0} & \textbf{4.0} & \textbf{5.0} &\textbf{6.0}&\textbf{7.0}  &\textbf{8.0}&\textbf{9.0}&\textbf{10.0} &\textbf{11.0}  & \textbf{12.0} &\textbf{13.0} & \textbf{14.0} & \textbf{15.0} &\textbf{16.0}& \textbf{17.0}  &\textbf{18.0} &\textbf{19.0}&\textbf{20.0}\\
		\hline\hline
			\multirow{2}{*}{NN}&	$\beta$ & 0.6& 1.0& 1.2 & 1.3 & 1.2& 0.8 & 0.3 & 0.2 &0.1 &0.1 &0.0& 0.0& 0.0 & 0.0 & 0.0 & 0.0  & 0.0 & 0.0 &0.0 &0.0 &0.0 & 0.0 \\
		\cline{2-24}
		&$\sigma$ & 1.1& 1.3& 1.0 &0.7 & 0.8& 1.3 & 1.3 & 1.3 & 1.3 & 1.7 &1.6 &1.4 & 1.4 & 1.3 & 1.3& 1.2 & 1.2 & 1.0 & 0.9 & 0.9 &0.8 &0.8      \\
	\hline
    \hline
	\multirow{2}{*}{NNN}&	$\beta$ & 0.4& 1.0& 1.1,$\ $1.2 & 1.3 & 1.2& 0.7 & 0.2 & 0.1 &0.1 &0.0 &0.0&  0.0& 0.0 & 0.0 & 0.0 & 0.0 & 0.0 & 0.0 &0.0 &0.0 &0.0 & 0.0 \\
	\cline{2-24}
	&$\sigma$ &3.1& 1.5& 1.1 & 1.0 & 1.2& 2.9 & 1.9 & 1.6 & 2.6 & 2.8 &2.6 &2.4& 2.4 & 2.3 & 2.2& 2.1 & 2.0 & 1.9 & 1.7 & 1.6 &1.4 &1.3   \\
	\hline
	\end{tabular}
	\label{Tab.new}
\end{table*}

\subsection{Integrability breaking at intermediate to large disorder $\gamma$: from 2d Poisson to class AI$^\dag$ and back}
\label{Sec:Ibreak}

In this section, we discuss the transition from random matrix statistics class AI$^\dag$ to 2d Poisson statistics, witnessed for intermediate ($\gamma=0.5$) to large ($\gamma=20$) disorder dissipation strength. The approach towards fully chaotic statistics represented by random matrix class AI$^\dag$ when increasing system size $L=12,14$ and 16 is included in subsection \ref{Sec:Ldependence} below. 

The smallest value of disorder for which a meaningful fit with a 2dCG 
at $\beta\geq0$ can be made is $\gamma=0.5$, when, $\beta=0.6\, (0.4)$ for NN (NNN), see Fig. \ref{Fig:intermedg} left. At this disorder strength, both Hermiticity and integrability appear to be broken. The system is in between the integrable case ($\beta=0$) and the fully chaotic case ($\beta\approx1.3-1.4$, equivalent to random matrix symmetry class AI$^\dag$). The latter is then attained at $\gamma=2$, where the maximal value for $\beta$ (and height of the maximum) is reached, see Figure \ref{fig:summary} right. We compare here the NN and NNN spacings with the 2dCG 
rather than AI$^\dagger$. This is because the AI$^\dag$ class is 
very well approximated by the 2dCG at these values of $\beta$, as shown in Fig. \ref{Fig:AI+} and Table \ref{Tab:NN and NNN AI dag 2dC}. Furthermore, random matrix statistics is approached in $\beta$ from below, see also the next subsection \ref{Sec:Ldependence}, and the $\sigma$-value gives a measure of how close it is.

Increasing $\gamma$ further leads to a rapid decrease in the best fitted value for $\beta$, see also Table \ref{Tab.new} below.  
From $\gamma \simeq 8-10$ onwards we consistently obtain $\beta=0$. This is because the disorder is so strong that the spins align with the disorder term and the statistics becomes again deterministic. 
The best fits for $\beta$ for NN and NNN and corresponding values, including standard deviations, are collected in Table \ref{Tab.new}. In this section we only fit to $\beta$ in the 2dCG  
where the step width is 0.1, but both NN and NNN distribution are available. The next section \ref{Sec:AsymptPoisson} presents a much finer fit in $\beta$ for $\gamma\geq 8$, possible for the NN only thanks to the surmise \eqref{Ps_2dCG}, which is a very good approximation so close to $\beta=0$. The approach to zero can also be seen well in the summary Fig.~\ref{fig:summary} right.  
Similar findings for Hermitian Hamiltonians comparing to 1d Poisson and corresponding classical random matrices, using both NN and NNN spacings, were made very recently  \cite{kundu2024beyond}.

The findings in this section correspond to the expected behaviour of a quantum chaotic system with imaginary disorder, where the spectral statistics lies between 2d Poisson and the corresponding random matrix symmetry class, in accordance with the GHS conjecture.

\begin{figure}[h!]
	\hspace{10pt}{$L=12$} \hspace{50pt}{$L=14$}\hspace{50pt}{$L=16$} \hfill\\
  
    \begin{turn}{90}\hspace{15pt}{NN}\end{turn}	
    	\includegraphics[width=0.15\textwidth,angle=0]{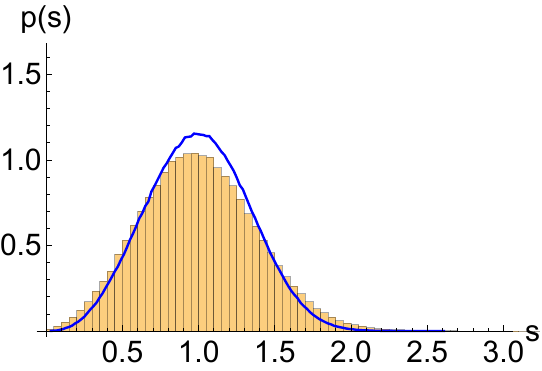}
	\includegraphics[width=0.15\textwidth,angle=0]{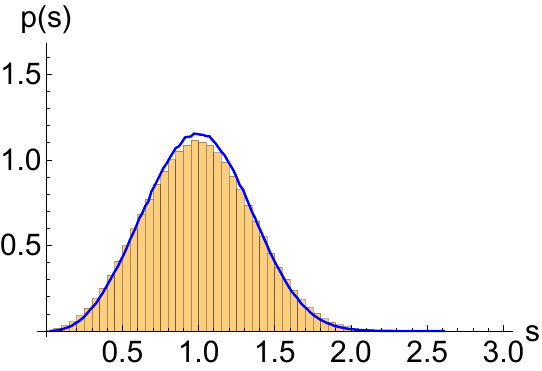}
	\includegraphics[width=0.15\textwidth,angle=0]{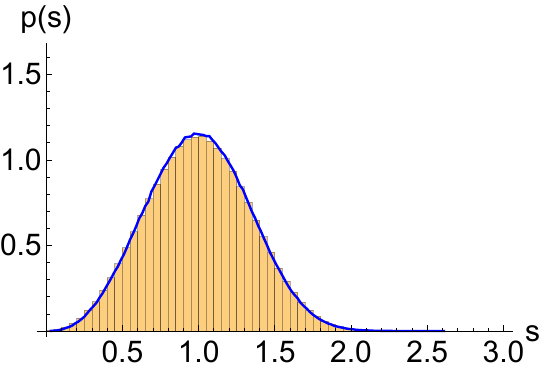}\\
  \hspace{10pt}{$\sigma=5.6\cdot 10^{-2}$} \hspace{30pt}{$\sigma=2.1\cdot 10^{-2}$}\hspace{30pt}{$\sigma=0.7\cdot 10^{-2}$} \hfill\\
  
    \begin{turn}{90}\hspace{15pt}{NNN}\end{turn}    
    \includegraphics[width=0.15\textwidth,angle=0]{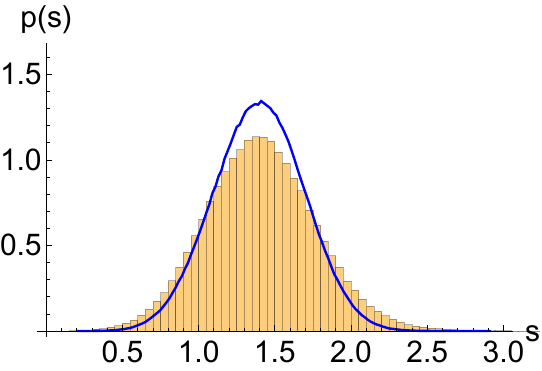}
    \includegraphics[width=0.15\textwidth,angle=0]{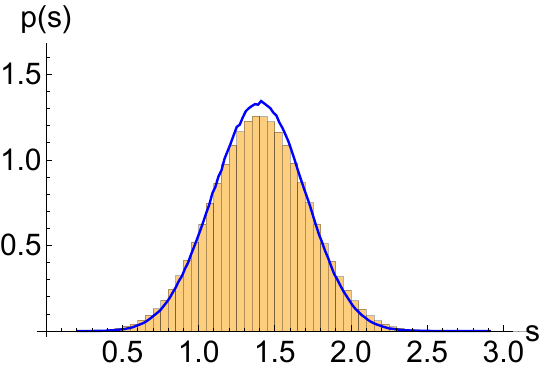}
    \includegraphics[width=0.15\textwidth,angle=0]{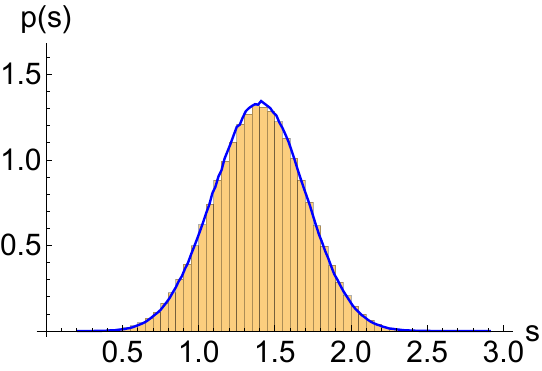}\\
 \hspace{10pt}{$\sigma=8.4\cdot 10^{-2}$} \hspace{30pt}{$\sigma=3.2\cdot 10^{-2}$ }\hspace{30pt}{$\sigma=1.1\cdot 10^{-2}$} \hfill\\
    \caption{
    Comparison of the XXZ with dissipative order $\gamma=2$ in the $\mathcal{S}_z = 0$ magnetization sector with increasing number of spins $L=12,14,16$ (from left to right, histograms) and the symmetry class AI$^\dagger$ of complex symmetric matrices (blue curve) of size $N=5000$. The NN distribution is given in the first row, whereas the NNN spacing is shown in the second row. 
    Additionally, we give the standard deviation $\sigma$ between the two plotted distributions below each plot. We observe a decrease of the standard deviation for growing $L$, as expected.}
    \label{Fig:L-g=2}
\end{figure}

\begin{figure}[]
    \centering
    \includegraphics{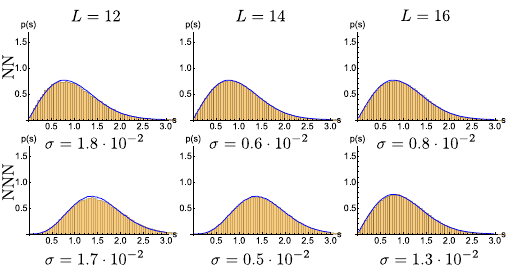}
    \caption{Same comparison as in Fig.~\ref{Fig:L-g=2} but at $\gamma=20$ (histograms), where all three system sizes agree with 2d Poisson statistics (blue curves). Here, $\sigma$ is comparable for all 3 systems sizes.
    }
    \label{Fig:L-g=20}
\end{figure}

\subsubsection{Dependence on system size $L$}
\label{Sec:Ldependence}

The main data presented so far are always for the longest spin chain, with $L=16$. This corresponds to a Hilbert space dimension of $12870$, that allows reaching a fully chaotic behaviour. 
Here, we also present data for system sizes $L=14$ and $L=12$, with respective dimensions of Hilbert space $3432$ and $924$, at the values $\gamma=2$ and $\gamma=20$. While for all three system sizes the 2d Poisson limit is reached fully at large $\gamma=20$, see Fig. \ref{Fig:L-g=20},  
this is not the case for the limit towards random matrix statistics of class AI$^\dag$, see Fig. \ref{Fig:L-g=2}.

At the very beginning of this section we have reminded the reader that in the $\beta$-range between 2d Poisson and class AI$^\dag$, the latter has the highest maximum for the NN spacing at the highest value of $\beta\approx 1.3-1.4$. In all three systems sizes, we observed that the highest maximum is obtained for disorder strength $\gamma=2$, 
thus being the closest to class AI$^\dag$. This is why, to look at the dependence on the system size $L$, we can focus on a single disorder strength. Figure \ref{Fig:L-g=2} shows that the best fit for $\beta$ at $\gamma=2$ increases with system size, and only reaches $\beta=1.3$ for the largest systems. This is summarised in Table \ref{Tab:L-g=2}. 
Notice that the maximal disorder found at $\gamma=2$ is not apparently related to any symmetry of the Hamiltonian.

\begin{table}[h]\centering
	\caption[]{Best fits for $\beta$ at the same disorder strength $\gamma=2$ for different systems sizes $L$, with standard deviation $\sigma$ in units $ 10^{-2}$.}\vspace{3mm}
	\begin{tabular}{|c|c|c|c|c|}
		\hline
		size &$L$  & 12 & 14 & 16 \\
		\hline
  \hline
		\multirow{2}{*}{NN}&	$\beta$ & 0.9 & 1.2 & 1.3   \\
		\cline{2-5}
		&$\sigma$ & 1.3 & 0.6 & 0.7 \\
		\hline
		\hline
		\multirow{2}{*}{NNN}&$\beta $& 0.8 & 1.1 & 1.3\\
		\cline{2-5}
		&$\sigma$ & 3.5 & 1.3 & 1.0\\
		\hline
	\end{tabular}
	\label{Tab:L-g=2}
\end{table}

\begin{table*}
    \centering 
	\caption[]{List of best fit $\beta$ for the comparison of the nearest-neighbour (NN) spacing of the spin chain with $L=16$ and $\gamma=8,\ldots 20$ to the NN 2dCG surmise \eqref{Ps_2dCG} and the corresponding standard deviation $\sigma$ in units of $10^{-2}$. }\vspace{3mm}
	\begin{tabular}{|c|c|c|c|c|c|c|c|c|c|c|c|c|c|c|}
		\hline
		&$\boldsymbol{\gamma}$ & \textbf{8} & \textbf{9} &\textbf{10}  & \textbf{11} &\textbf{12} & \textbf{13} & \textbf{14} &\textbf{15}& \textbf{16}  &\textbf{17}&\textbf{18}&\textbf{19} &\textbf{20} \\
		\hline\hline
			\multirow{2}{*}{NN} &	$\beta$ & 0.0547 & 0.0413 & 0.0337 & 0.0283 & 0.0273 & 0.026 & 0.023 & 0.0196 &
       0.0169 & 0.0135 & 0.0108 & 0.0066 & 0.0035 \\
		\cline{2-15}
		&$\sigma$ & 0.36 & 0.35 & 0.33 & 0.35 & 0.33 & 0.33 & 0.31 & 0.3 & 0.27 & 0.24 & 0.24 &
       0.21 & 0.2      \\
	\hline

	\end{tabular}
	\label{Tab:g8-20}
\end{table*}

\begin{figure*}
    \centering
    \includegraphics[width=\linewidth]{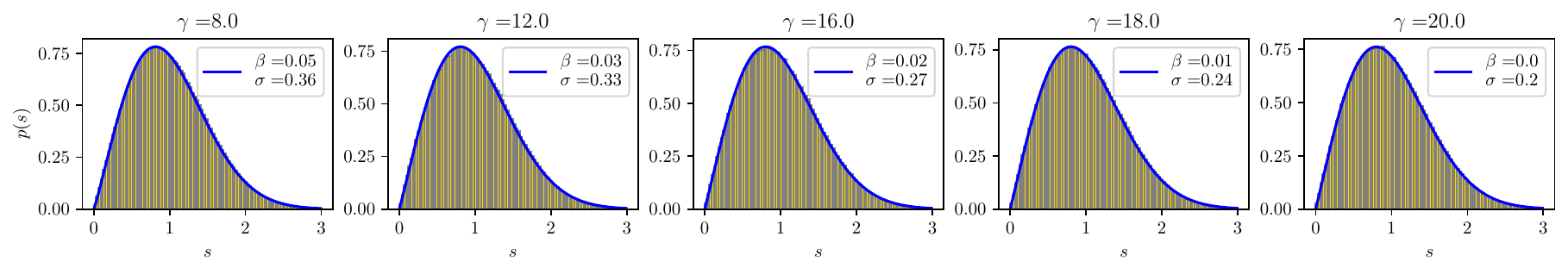}
    \caption{Asymptotic approach to 2d Poisson for large disorder strength $\gamma$. The histograms for the NN spacings of at least 450 realizations of the spin chain with $L=16$ in the $\mathcal{S}_z = 0$ magnetization sector are compared with the best $\beta$ fit of the surmise \eqref{Ps_2dCG} shown as a smooth blue line. The legend shows the value of $\beta$ and the standard deviation (in units of $10^{-2}$) of the fit.}
    \label{fig:beta0_approach}
\end{figure*}

\subsection{Asymptotic approach to 2d Poisson for large disorder $\gamma$}
\label{Sec:AsymptPoisson}

This subsection details the asymptotic approach to 2d Poisson statistics $\beta\to0$ for large values of $\gamma=8-20$. Because for the 2dCG 
we have only discrete steps in  $\beta$ of size 0.1 available, we will use the surmise \eqref{Ps_2dCG} instead here. It is an excellent approximation for the local statistics close to $\beta=0$, but it only allows fitting the NN spacing distributions. Fig.~\ref{fig:beta0_approach} shows fits in $\beta$ of the surmise to the NN spacing distribution for various values of $\gamma=8-20$ in steps of size 2 (or more). 
The fitted $\beta$-values for all values of $\gamma\geq8$, together with the corresponding standard deviations, are provided in Table \ref{Tab:g8-20}.




\subsection{Comparison to complex spacing ratio discussion}
\label{Sec:XXZiW_CSR}
\vspace{-1ex}

\begin{figure*}[]
\includegraphics[width=\linewidth]{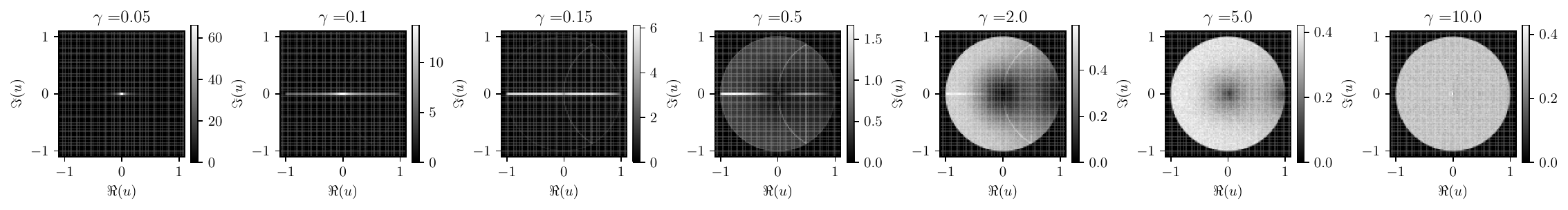}
 \caption{Complex spacing ratio histograms for the XXZ model with dissipative disorder \eqref{XXZ_iW_Hamiltonian} with $L=16$ spins in the $\mathcal{S}_z=0$ sector, for various values of disorder strength $\gamma$. For each value of $\gamma$, the data shown is for at least 450 realizations, in the bulk of the spectrum. }
 \label{fig:CSRs}
 \end{figure*}

The probability density of the complex spacing ratios, as seen in Figure~\ref{fig:CSRs}, contains interesting patterns especially in the chaotic regime: the lower density around the origin and around $r=1$ are due to level repulsion. It is not likely for an eigenvalue and its NN to be close together, thus the density is lower at $r=0$; and an eigenvalue's NN and NNN would most likely lie on either side of it, making the density lower at $r=1$; see also Figure \ref{Fig:ratios} which compares the complex spacing ratio distribution of 2d Poisson with that of AI$^\dagger$. 
The `bow and arrow' feature seen for small $\gamma$ is a manifestation of an underlying structure of the complex eigenvalues. As observed and explained in Appendix D of~\cite{Li:2024uzg}, this feature is a result of an eigenvalue's NN and NNN being on the real line and as complex conjugate to it, or vice versa. Section~\ref{Sec:PH_spectrum} showed that the Hamiltonian is pseudo-Hermitian and thus its spectrum is symmetric to reflection across a line parallel to the real axis.  In the complex spacing ratios, this symmetry manifests itself as reflection symmetry across the real axis. Additionally, the excess of ratios along the real line is a result of real eigenvalues having real  NN and NNN. 

\section{Spectral statistics of complex eigenvalues}
\label{Sec:2d_Spectral}
In this section and in the next we describe the tools we have been using to analyze our results in the previous section.
In particular, we introduce and compare two different random point processes of $N$ points $z_1,\ldots,z_N$ in the complex plane (including the real line as a special case), namely the 
2dCG and the $D$-dimensional Poisson point process, in addition to the class AI$^\dag$, which can be well approximated by the 2dCG with $\beta=1.4$ \cite{akemann2022spacing}. These processes are characterised by a joint distribution $\mathcal{P}_N(z_1,\ldots,z_N)$. We will recall exact and approximate analytical expressions, as well as numerically generated distributions for the local spacing between complex eigenvalues. 

The NN is given for each point $z_i$ by finding the NN in radial (Euclidean) distance as $s_i^{\textsc{nn}}=\min_{j\neq i}|z_i-z_j|$, being realised say by $j = i_0$ \footnote{Such point may not be unique, but the event of finding two neighboring point at the very same distance will be of measure zero.}. 
The NNN is then at distance $s_i^{\textsc{nnn}}=\min_{j\neq i,i_0}|z_i-z_j|$. 
When determining the respective distributions numerically, we first normalise the mean density by unfolding, as described in Sec.~\ref{Sec:unfolding}, and then determine the NN and NNN for all $i=1,\ldots,N$ to obtain a distribution. We will be careful to only take points inside the so-called bulk of the spectrum, that are sufficiently away from the edge of the spectrum---at order $O(1/\sqrt{N})$ for the random matrix ensembles we study. In all our examples, the spectral density has compact support \footnote{It is expected based on numerical simulations that the spacing distribution at the edge is different
}.

 
\subsection{The 2-dimensional Coulomb gas and a surmise}
\label{Sec:2dCG}

The first example of a point process we compare our data with is the 2dCG, 
featuring a logarithmic interaction among the $N$ point charges, with a strength that can be interpreted as an inverse temperature $\beta$.  Choosing a Gaussian confining potential, the joint distribution in the complex plane reads:
\begin{equation}
    \label{JPD_2dCG}
    \!\mathcal{P}_{N,\beta}(z_1,z_2,\dots,z_N)\! \propto \!\exp\Big[{-}\sum_{i=1}^N |z_i|^2 {+}\frac{\beta}{2} \sum_{i\neq j}^N \ln|z_i-z_j|\Big].\!
\end{equation}
The normalisation constant is given by the complex Selberg integral, which can be solved  analytically only for $\beta=0$ or 2. 
Compared to the standard statistical mechanics notation, where the Gaussian term has a coefficient $\beta/2$, the positions of the point charges are rescaled as $z_i \to \sqrt{2/\beta}z_i$.  
This rescaling allows us to take the limit $\beta \to 0$ and thus to recover the 2d Poisson case \footnote{The effect of this rescaling on the log term amounts to adding a constant $(\beta/2)\ln \sqrt{2/\beta}$, which in turn becomes a multiplicative factor, $\exp[\frac{\beta N(N-1)}{4}\ln \sqrt{\beta/2}]$, in the joint distribution. In the limit $\beta\to0$, this factor goes to 1}. 
In addition, the three different local symmetry classes of non-Hermitian random matrices can be obtained: namely, Eq.~\eqref{JPD_2dCG} with $\beta=2$ corresponds to the complex eigenvalue distribution of the complex Ginibre ensemble (class A), which can be solved analytically, cf. Appendix \ref{App:RMT}; in turn, $\beta=1.4$ and $\beta=2.6$ give a very good approximation of the NN spacing distribution of class AI$^\dag$ and class AII$^\dag$, respectively \cite{akemann2022spacing}. Below, we will show that this approximation extends to NNN, too.
 To date, no analytic results for the local NN and NNN spacing are known apart for $\beta=0,2$ \footnote{For $N=2$ the spacing distribution is known for AI$^\dagger$ and AII$^\dagger$ \cite{hamazaki2020universality,jaiswal2019universality}, which are however not a good approximation for the large-$N$ limit as was discussed in \cite{akemann2022spacing}}.
The numerically determined spacings in the 2dCG are shown in Fig. \ref{Fig:2d-statistics}(b).

Interestingly, in the vicinity of $\beta=0$, the NN spacing distribution of the 2dCG 
 can be well approximated by a surmise  \cite{akemann2022spacing} based on $2\times2$ matrices. As a function of $\beta$ it is given by \footnote{There is a typo in equations (23) and (27) of reference~\cite{akemann2022spacing}, we write down the correct formula in~\eqref{Ps_2dCG}.}
\begin{eqnarray}
\label{Ps_2dCG}
    p_{\textsc{nn},\beta}^{\textrm{surmise}}(s) &=& \frac{2\,\alpha_{\textrm{eff}}^{1+\beta_{\textrm{eff}}/2}}{\Gamma(1+\beta_{\textrm{eff}}/2)}s^{1+\beta_{\textrm{eff}}} \exp[-\alpha_{\textrm{eff}}\,s^2]\\
        \alpha_{\textrm{eff}} &=& \frac{\Gamma^2[(3+\beta_{\textrm{eff}})/2]}{\Gamma^2(1+\beta_{\textrm{eff}}/2)} ~,
    \nonumber
        \end{eqnarray} 
where 
\begin{equation}
    \beta_{\textrm{eff}}(\beta) = 2.108\beta - 0.19 \beta^2 +0.03 \beta^3.
\end{equation}
This effective beta is introduced to improve the approximation to the true  
2dCG up to approximately $\beta=0.5$, see the discussion in  \cite{akemann2022spacing}. 
Setting $\beta=0$ in~\eqref{Ps_2dCG} exactly reproduces the NN spacing distribution \eqref{PoiD-NN} for 2d Poisson.
The distribution $p_{\textsc{nn}, \beta}^{\textrm{surmise}}(s)$ and its first moment are again normalised to unity. 
Even at larger values of $\beta$, this surmise, while approximate, can be very useful to do a first fit of $\beta$ that can then be improved by finding the best fit to the (numerically generated) spacings of the true 2dCG. 

We note in passing that when restricting all points in \eqref{JPD_2dCG} to the real line, the joint density of real eigenvalues is that of the common  Hermitian Gaussian random matrix ensembles, GOE, GUE and GSE, at respective values $\beta=1,2,4$ (see e.g.~\cite{mehta2004random}). This is also known as a Dyson gas.
In addition to these values, where the ensemble can be solved analytically and the statistics is well understood, the general case $\beta>0$ has been well studied \cite{forrester2010log} (and the normalising Selberg integral is then known for any $N$ and $\beta$). In this Hermitian setup, a popular tool is the Wigner surmise based on $2\times2$ matrices, approximating the local NN spacing distribution well for $\beta=1-4$ away from 1d Poisson at $\beta=0$. 
All 3 Hermitian ensembles have distinct local statistics and exhaust the 3 random matrix bulk statistic classes for real eigenvalues. 
This is in stark contrast to the three non-Hermitian Ginibre ensembles with real,  complex and quaternion elements, since these three share the same local statistics in the bulk of the spectrum away from the real line, as shown analytically in \cite{akemann2019universal, borodin2009ginibre}.


\subsection{The D-dimensional Poisson Point Process}
\label{Sec:DPoisson}

The second point process we compare our data to is the Poisson point process in $D$ dimensions, which has been much studied in the literature. 
It is given by $N$ points distributed randomly and independently in a $D$-dimensional ball of radius $R$, keeping the density of points fixed by setting $N/R^D=1$. 
The joint distribution $\mathcal{P}_N(z_1,\ldots,z_N)$ for Gaussian random variables thus corresponds to \eqref{JPD_2dCG} at $\beta=0$, and completely factorises. 
After appropriate rescaling, the NN spacing distributions in $D=2$ \cite{haake1991quantum} and  for general $D$ \cite{sa2020complex} can be derived analytically in the large-$N$ limit, yielding
\begin{equation}
\label{PoiD-NN}
    P^{\textsc{nn}}_{\textrm{Pois},D}(s) = D\,\Gamma\Big(1+\frac{1}{D}\Big)^D s^{D-1} \, e^{-\Gamma\left(1+\frac{1}{D}\right)^D s^D}.
\end{equation}
At $D=1$, it recovers the familiar exponential distribution $P^{\textsc{nn}}_{\textrm{Pois},1}(s) =e^{-s}$, whereas at $D=2$, it is Gaussian with a linear prefactor stemming from the area measure, see also \cite{sakhr2006wigner}. 
In turn, the NNN spacing distribution can be derived by integrating over the joint distribution of NN and NNN, provided in e.g.~\cite{sa2020complex}, to find
\begin{equation}
\label{PoiD-NNN}
    P^{\textsc{nnn}}_{\textrm{Pois},D}(s) = D\,\Gamma\Big(1+\frac{1}{D}\Big)^{2D} s^{2D-1} \, e^{-\Gamma\left(1+\frac{1}{D}\right)^D s^D}.
\end{equation}
Both distributions are normalized to unity. The NN distribution is rescaled to have its first moment equal to one, and the NNN distribution is rescaled by the very same factor. 
Based on the BT conjecture, generic integrable Hamiltonian systems are expected to display Poisson statistics at $D=1$ when being Hermitian, and $D=2$ Poisson when
non-Hermitian, see e.g. \cite{akemann2019universal,sa2020complex}. Because of this, we assume that both 2d Poisson NN {\it and} NNN spacings describe spectral data of such Hamiltonians. This was numerically confirmed very recently for the BT conjecture in 1d in \cite{kundu2024beyond}. 

Since both distributions \eqref{PoiD-NN} and \eqref{PoiD-NNN} are analytic in $D$, we can continue them to arbitrary $D>0$, as illustrated in Fig.~\ref{Fig:2d-statistics}(a). 
Such a Poisson distribution with an effective dimension $1<D<2$ was observed e.g. on forest patches,  cf.~\cite{Akemann_2024}.


\subsection{Non-Hermitian random matrix class AI$^\dag$}
\label{Sec:AIdagger}

According to the GHS conjecture, fully chaotic non-Hermitian Hamiltonians follow local non-Hermitian random matrix statistics; therefore, we have to compare our model \eqref{XXZ_iW_Hamiltonian} to the random matrix ensemble corresponding to the same symmetry class. In Ref.~\cite{hamazaki2020universality}, it was conjectured, based on numerical simulations, that among all 38 classes of non-Hermitian random matrices, there are only three different classes of local level spacing statistics in the bulk of the spectrum. Following their notation, the representatives of these 3 different symmetry classes are given by the complex Ginibre ensemble (class A), Gaussian complex symmetric (class AI$^\dag$), and complex self-dual 
(class AII$^\dag$) matrices. The joint density is known only for class A, then given by \eqref{JPD_2dCG} with $\beta=2$.

\begin{figure}[h!]
		\centering
		\includegraphics{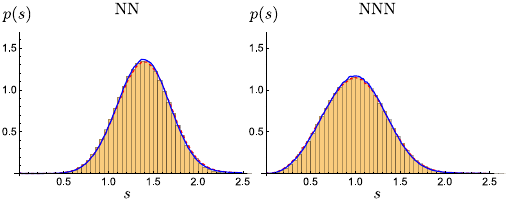}
		\caption{The NN (left) and NNN (right)  AI$^\dag$ spacing distribution generated numerically (histograms) for 500 random matrices of size $N=5000$. Also shown are comparisons to the NN and NNN spacing distribution of the 2dCG, 
  with fitted values $\beta=1.3$ (red dashed) and $\beta=1.4$ (blue full curve).  The comparison with NNN is new here. 
}
		\label{Fig:AI+}
	\end{figure}

Since our Hamiltonian \eqref{XXZ_iW_Hamiltonian} is also complex symmetric, as shown in Section \ref{Sec:model}, we compare the NN and NNN spacing distributions with those of random matrices in class AI$^\dag$. Such matrices are defined by symmetric $N\times N$ matrices $J$ with complex elements and  Gaussian distribution:
\begin{equation}
    J = J^T, \quad P_N^{(\rm{AI}^\dag)}(J)=C_N e^{-\textrm{Tr}(JJ^\dagger)}, 
\end{equation}
with normalisation $C_N^{-1}=\pi^N(\pi/2)^{N(N-1)/2}$. Apart from $N=2$ \cite{hamazaki2020universality,jaiswal2019universality}
and the local limiting spectral density of rescaled imaginary parts (weak non-Hermiticity)  
\cite{Sommers_1999}, 
 there are no analytic results available for local statistics in this class so far. 
 While $N=2$ is a good approximation for the Wigner surmise for Hermitian random matrices, it does not work here \cite{akemann2022spacing}. So we have to generate the spacing distributions numerically, see Fig.~\ref{Fig:AI+}.

It was found in \cite{akemann2022spacing} that the NN spacing of class AI$^\dag$ distribution can be very well fitted by a 2dCG  
at $\beta=1.4$, which is also numerically generated, see the next subsection \ref{Sec:2dCG}. In this work, we also compare to the NNN spacing of AI$^\dag$, at the same value of $\beta=1.4$, and find an excellent agreement, see Fig. \ref{Fig:AI+} right. 
The quality of the best fit for $\beta$ (marked in bold) is shown in standard deviations $\sigma$ \eqref{def-sigma} and Kolmogorov-Smirnov distances $d$ \eqref{def-KS} { in Table \ref{Tab:NN and NNN AI dag 2dC}}, to be introduced in Subsection \ref{Sec:numerical}. The results for class A are presented in Table \ref{Tab:NN and NNN A 2dC}.

These results further corroborate that the AI$^\dag$ random matrix symmetry class is very well approximated by a 2dCG. 
The same conclusion applies to the NNN spacing distribution of class AII$^\dag$, shown in Appendix \ref{App:RMT}, where we also discuss the quality of the fit comparing numerically generated spacings from class A with its analytic results.

\begin{table}[h!]\centering
	\caption[]{List of standard deviations, $\sigma$, and Kolmogorov-Smirnov distances, $d$, (both in units $ 10^{-2}$) 
		between NN and NNN spacings of AI$^\dagger$ and 2dCG,  
		with $\beta\in[1.1,1.6]$ in steps of $0.1$. For NN, both measures favour $\beta=1.4$ as best fit, as does the Kolmogorov-Smirnov distance for NNN, 	
		 whereas 
		the standard deviation for the latter is best for $\beta=1.3$ (bold numbers).}.\vspace{3mm}
	\begin{tabular}{|c|c|c|c|c|c|c|c|}
		\hline
	fitted	&$\beta$  & 1.1 & 1.2 & 1.3 & 1.4 & 1.5& 1.6\\
		\hline
  \hline
	\multirow{2}{*}{NN}&	$\sigma$ & 3.3 & 2.2 & 1.1 & \bf{0.9} & 1.7 & 2.7 \\
		\cline{2-8}
		&$d$ & 1.7 & 1.2 & 0.6 & \bf{0.3} & 0.7 & 1.2 \\
		\hline
		\hline
		\multirow{2}{*}{NNN}&$\sigma $& 3.4 & 2.0 & \bf{0.8} & 1.4 & 2.7 & 4.0 \\
		\cline{2-8}
		&$d$ & 3.0 & 1.8 & 0.8 & \bf{0.6} & 1.7 & 2.7\\
		\hline
	\end{tabular}
	\label{Tab:NN and NNN AI dag 2dC}
\end{table}


\section{Methods}\label{Sec:methods}
This section presents the methods we use to study and analyse spectral statistics of the physical model, including unfolding, 
complex spacing ratio statistics and fitting procedures.

\subsection{Unfolding complex spectra}
\label{Sec:unfolding}

When comparing data from physics (or other sciences), e.g.\ taken by discrete (energy) eigenvalues $E$ of Hermitian or non-Hermitian Hamiltonians, to universal predictions e.g.\ from Poisson or random matrix statistics, some pre-processing is needed. Typically, the mean spectral density $\bar{\rho}(E)$ of the physical system contains model specific (and physical) information.
In order to have a chance to see universal statistics, one has to separate scales in the density $\rho(E)$ between the mean density and the fluctuation (fl) around the mean (also called local or microscopic):
\begin{equation}
    \rho(E) = \bar{\rho}(E) + \rho_{\mathrm{fl}}(E) ~.
\end{equation}
For geometric reasons, with $N$ the number of points and $\bar{\rho}(E)\sim N$, these fluctuations are typically of the order of $1/N$ in 1d, and of order $1/\sqrt{N}$ in 2d, at least in the bulk of the spectrum. 

The removal of the mean density, or in other words, map to a normalised constant density, is called unfolding. In 1d, this is a unique and well understood procedure, see e.g. \cite{guhr1998random}. In 2d however, there is more freedom and several different methods have been proposed, e.g.~\cite{markum1999non, akemann2019universal, hamazaki2020universality, suthar2022non, garcia2023universality, Li:2024uzg}. We follow the procedure described in~\cite{akemann2019universal}, which we briefly recall below.

We start from the exact density of states of $N$ eigenvalues $z_{i=1,\ldots, N}$ in 2d or the complex plane given by 
\begin{equation}
    \rho(x,y) = \frac{1}{N}\sum_{i=1}^N \delta^{(2)} (z-z_i),
\end{equation}
where $z=x+iy$.
The mean, or average, density of states can be approximated by a smooth function by replacing each delta function with a Gaussian distribution with a certain variance $\Sigma$: 
\begin{equation}
    \bar{\rho}(x,y) = \frac{1}{2\pi \Sigma^2 N} \sum_{i=1}^N e^{-\frac{|z-z_i|^2}{2\Sigma^2}}. 
\end{equation}
Here, $\Sigma= \alpha \bar{s}$ is determined by multiplying the average NN spacing of the spectrum, $\bar{s}$, by a constant $\alpha$ of $O(1)$ in order to smear over several eigenvalues on average  
\footnote{In~\cite{akemann2019universal} it was suggested that $\alpha=4.5$ is a good choice and we use this value too.}.
The unfolded NN and NNN spacings $s_i'^{\textsc{nn}}$ and  $s_i'^{\textsc{nnn}}$ are then defined by removing the local scale, that is, by multiplying the original spacings by the same factor 
\begin{subequations}
\begin{align}
s_i'^{\textsc{nn}} &= |z_i - z_i^{\textsc{nn}}|\sqrt{\bar{\rho}(x_i,y_i)},
\\
s_i'^{\textsc{nnn}} &= |z_i - z_i^{\textsc{nnn}}|\sqrt{\bar{\rho}(x_i,y_i)},
\end{align}
\end{subequations}
where $z_i^{\textsc{nn}}$ is the NN to eigenvalue $z_i$, and $z_i^{\textsc{nnn}}$ the NNN. 
After unfolding, we normalise the NN spacing such that the first moment is equal to one and measure the NNN spacing on the same scale.


\subsection{Complex spacing ratios}
\label{Sec:CSR}

\begin{figure}[h]
		\centering
		\includegraphics[width=0.49\linewidth,angle=0]{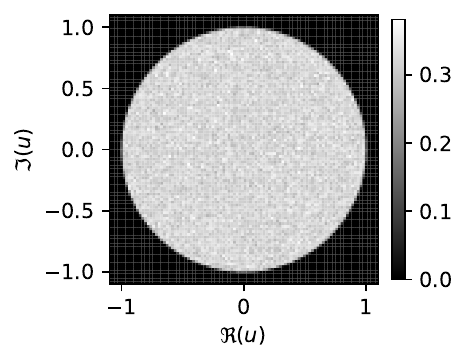}
		\includegraphics[width=0.49\linewidth,angle=0]{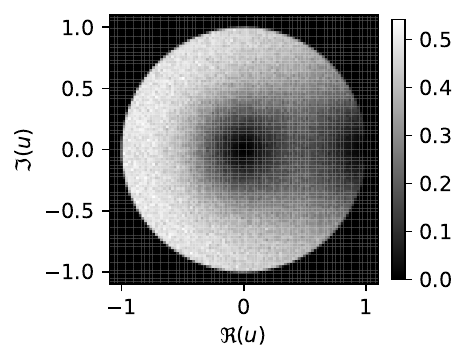}
		\caption{Numerically generated complex spacing ratio distribution $R(u=re^{i\theta})$ for 2d Poisson (left) and random matrices in class AI$^\dag$ (right). Both plots are based on size $N=5000$, with 500 realisations.
		Class AI$^\dag$ looks qualitatively similar to class A, with a pronounced hole at the origin and depletion from unity, see \cite{sa2020complex,kanazawa2021new}.}
		\label{Fig:ratios}
	\end{figure}

Alternatively to the NN (and NNN) spacing distribution, a tool that does not require unfolding relies on the spacing ratios, as proposed in 1d \cite{Oganesyan2007Localization}
 and 2d \cite{sa2020complex}. In 2d, the complex spacing ratio is defined as 
\begin{equation}
\label{2dratio-def}
   u_i = \frac{z_i^{\textsc{nn}}-z_i}{z_i^{\textsc{nnn}}-z_i}~,
\end{equation}
and the goal is to provide (analytical) predictions for the limiting probability distribution $R(u=re^{i\theta})$ as a function of radius $r$ and angle $\theta$. For example, the 2d Poisson process has a flat distribution over the unit disc, $R_{\rm Pois}(u)=\Theta(1-|u|)/\pi$, as shown in Fig. \ref{Fig:ratios} left.
In view of the difficulties to make 2d fits to such ratios, the radial and angular distribution were also introduced, respectively by averaging  $R(u=re^{i\theta})$ over angle $\theta$ or radius $r$. Moments were computed for the complex Ginibre ensemble class A \cite{sa2020complex} and for classes AI$^\dag$ and AII$^\dag$ \cite{kanazawa2021new}. 

In 1d, the joint density and its normalisation are well understood. This has led to a plethora of predictions for limiting ratios, cf. \cite{atas2013distribution}, including for further neighbours and, most importantly, for their dependence on the inverse temperature $\beta$ in the Dyson gas.  

In contrast, in 2d the only approximate results for radial and angular moments are known for class A \cite{dusa2022approximation}, which is not the  symmetry class of our model. For the random matrix class AI$^\dag$ applicable here, complex spacing ratios \eqref{2dratio-def} can be computed numerically, see Fig. \ref{Fig:ratios} right, as well as the corresponding radial and angular moments \cite{kanazawa2021new}. However, it is not clear how the transition between 2d Poisson and AI$^\dag$ would look like. 
Moreover, in the analysis of data from ecology \cite{Akemann_2024}, the behaviour of radial and angular moments with effective $\beta$  between the 2d Poisson and class A random matrix predictions were found to be inconclusive. For this reason, we did not pursue a quantitative comparison between our data and complex spacing ratios.

\subsection{Comparison of numerical data}
\label{Sec:numerical}
In order to compare the spectral distributions of the spin chain model with dissipative disorder to those from Poisson, 2dCG, 
or random matrix of class AI$^\dagger$, we present two measures of agreement between two distributions $p_1(s)$, $p_2(s)$. Both measures are based on numerical data. The first one is the \textit{standard deviation} $\sigma$, defined as
\begin{align}
\label{def-sigma}
    \sigma=\bigg[\frac{1}{n}\sum_{j=1}^n [p_1(s_j)-p_2(s_j)]^2\bigg]^{\frac{1}{2}},
\end{align}
where $n$ is the number of bins in the data and $p_1(s_j)$, respectively $p_2(s_j)$, are the number of counts in the $j$-th bin at its mid point $s_j$ in each distribution. Hence, one first determines a fine binning of the data into a histogram, and then uses these histogram points $(s_j)_{j=1}^n$ for the comparison. 
The sum was cut off at $s = 3$ when fitting to the 2dCG, because of the exponential suppression; and at a larger $s=7$ when fitting to a generalized $D$-dimensional Poisson process.

The second measure is the so-called \textit{Kolmogorov-Smirnov distance} $d$, which is defined as
\begin{align}
\label{def-KS}
    d=\max_{x\geq 0}|F_1(x)-F_2(x)|\in [0,1],
\end{align}
between the cumulative distributions $F_1$ and $F_2$ of distributions $p_1$ and $p_2$, respectively. It is independent of the binning into histograms. However, the calculations take much longer than the ones for the standard deviation. We checked that the Kolmogorov-Smirnov distances and the standard deviations have the same sensitivity to, e.g.
$\beta$ for the comparison of the spin chain data and the 2dCG.

As a reference, we generated complex eigenvalues from class A (of size $N = 5000$), even though it exactly corresponds to the 2dCG with $\beta=2$. The density of class A and the 2dCG 
is flat for large-$N$, and thus no unfolding is needed~\cite{haake1991quantum,markum1999non,akemann2019universal}.  
Such fits, presented in Table \ref{Tab:NN and NNN A 2dC}, serve as a comparison to judge the quality of the other best fits
for matrices of this size in the other two classes, AI$^\dag$ and AII$^\dag$ \footnote{It appears from numerics that AI$^\dag$ and AII$^\dag$ also have a flat spectrum in the bulk and thus do not require unfolding.}. 

\begin{table}[]\centering
	\caption[]{Standard deviations $\sigma$ and Kolmogorov-Smirnov distances $d$ (both in units $ 10^{-2}$) 
		between numerically generated NN and NNN spacings of class A and the numerically generated 2dCG at $\beta=2$.} \vspace{3mm}
	\begin{tabular}{|c|c|c|}
		\hline
	$\beta=2$  & NN & NNN \\
		\hline
  \hline
  $\sigma$ & 0.44 & 1.4\\
  \hline
  $d$ & 0.2 & 1.2 \\
  \hline
	\end{tabular}
	\label{Tab:NN and NNN A 2dC}
\end{table}

The values obtained for $\sigma$ and $d$  for class AI$^\dag$, Table \ref{Tab:NN and NNN AI dag 2dC}, and  for class AII$^\dag$, Table \ref{Tab:NN and NNN AII dag 2dC}, are comparable to that of class A (which we know exactly).
For the comparison, we used the library of spacings in the 2dCG 
that was generated numerically in~\cite{akemann2019universal} with $N=200$ point charges. 
It was subsequently increased to $N=5000$ in \cite{akemann2022spacing} to improve the statistics of our Coulomb data, in particular  in the vicinity of $\beta=0$. Note that the best fit for the NNN is systematically less good than for NN data. This could come from the maxima  
of the NNN curves, which are always larger than the NN curves. The maxima of the distribution being larger, they
could sum into larger deviations, and thus yield a larger maximal difference for the cumulative distribution function.


\section{Conclusions}
\label{Sec:concs}
Using the NN and NNN distributions, this work shows that the XXZ model with dissipative disorder exhibits two spectral crossovers as a function of disorder strength $\gamma$.  At $\gamma=0$, the model is a pure integrable XXZ spin chain and its spectral statistics (once restricted to an appropriate symmetry sector to avoid exact degeneracies) corresponds to the 1d Poissonian statistics.  As the dissipative disorder strength is increased, but still small, the spectrum begins spreading into the complex plane and the statistics, 
approximately still integrable, are close to Poisson with an effective dimension $D>1$.  As the dissipative disorder is increased further, integrability begins to break and both the NN and NNN distributions are well fitted by the 2dCG with $\beta>0$. At $\gamma=2$, the model becomes fully chaotic with NN and NNN spectral statistics close to those of random matrices of class AI$^\dagger$, which is in the same non-Hermitian symmetry class as our model. Beyond $\gamma>2$,
we observe a crossover from chaos to integrability.  The same crossover from chaos to integrability was detected in~\cite{roccati2024diagnosing} through real singular value statistics and is now validated also from complex spectral analysis. In a similar manner to the Hermitian case, increasing the dissipative disorder induces a crossover of the spectral statistics from chaotic---described by random matrix symmetry class AI$^\dag$ or  the 2dCG statistics with $\beta\approx 1.3-1.4$, to integrable---described by the statistics of the 2d Poisson point process (or equivalently by the 2dCG 
with $\beta=0$).  We tracked the crossover by fitting the NN and NNN spectral distributions to the corresponding value of $\beta$ in the 2dCG, 
obtaining a consistent value for $\beta$ from both fits. We complemented by studying also the complex spectral ratios as a function of the disorder.

\acknowledgments
We acknowledge funding from the John Templeton Foundation (JTF Grant 62171) and the Luxembourg National Research Fund (FNR, Attract grant 15382998).  FR acknowledges financial support from the Fulbright Research Scholar Program. The opinions expressed in this publication are those of the authors and do not necessarily reflect the views of the JTF. The numerical simulations presented in this work were in part carried out using the HPC facilities of the University of Luxembourg.
This work was partly supported by the Deutsche Forschungsgemeinschaft (DFG) grant SFB 1283/2 2021–317210226 (GA, PP) and a Leverhulme Trust Visiting Professorship, grant VP1-2023-007 (GA). GA is indebted to the School of Mathematics, University of Bristol, where this research was conducted. AC acknowledges the CNLS at Los Alamos National Laboratory, where part of this research was conducted. 

\appendix

\renewcommand\thefigure{\thesection\arabic{figure}}    
\setcounter{figure}{0}    

\section{Random matrix bulk symmetry classes}\label{App:RMT}

This appendix briefly recalls what is known about local statistic in non-Hermitian random matrix theory. We focus on the 3 symmetry classes that are conjectured to exhaust the local bulk statistics \cite{hamazaki2020universality}. Representatives for these 3 bulk symmetry classes are class A, AI$^\dag$ and AII$^\dag$. 
We present NN and NNN spacing distributions and compare to the 2dCG, the class AI$^\dag$ being analysed 
in Subsection \ref{Sec:AIdagger}. 
Our new results include a comparison of the respective NNN with a fit to the 2dCG with the \textit{same} $\beta$,
for class AI$^\dag$ in Fig. \ref{Fig:AI+} (right), and for  AII$^\dag$ in Fig. \ref{Fig:AII+} (right). 
For illustration, we also compare the NN and NNN distribution of all 3 random matrix classes in Fig. \ref{Fig:3RMTs} below. 
Note that, based on pertubation theory, it was shown in \cite{grobe1989universality,haake1991quantum} that for small spacings the repulsion is cubic, $s\sim s^3$, irrespectively of the presence or absence of time reversal symmetry.

	\begin{figure}[b!]
		\centering		
		\includegraphics{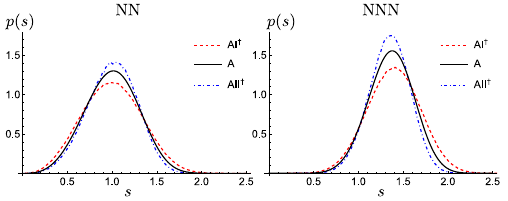}
		\caption{The NN (left) and NNN (right)  spacing distributions of the 3 random matrix  symmetry classes. Namely, in order of increasing maxima, class AI$^\dag$ (red dashed line), class A (black full line), and class AII$^\dag$ (blue dashed-dotted line).  These respectively correspond to an increasing $\beta\approx 1.3-1.4$, $\beta=2$, and $\beta\approx2.5-2.6$.
}
		\label{Fig:3RMTs}
	\end{figure}

\subsection*{Class A - definition and analytic results}
The simplest and best understood is class A, the complex Ginibre ensemble, consisting of complex non-Hermitian $N\times N$ random matrices $J\neq J^\dag$ with independent Gaussian elements and no further symmetry: 
\begin{equation}
\label{PA}
P_N^{(\rm{A})}(J)=C_N e^{-\textrm{Tr}(JJ^\dagger)}, \quad C_N^{-1}=\pi^{N^2}.
\end{equation}
Its joint distribution of complex eigenvalues exactly corresponds to the 2dCG \eqref{JPD_2dCG} at $\beta=2$. 
So it represents a determinental point process, and all complex eigenvalue correlation functions are known explicitly at finite- and large-$N$. 
The NN \cite{grobe1988quantum} and NNN \cite{akemann2009gap} spacing distributions in the large-$N$ limit read
\begin{eqnarray}\label{ANN}
	p_{\textsc{nn}}^{\rm (A)}(s) &=&
E^{\rm (A)}_{\textsc{nn}}(s)
	\sum_{j=1}^{\infty} \frac{2s^{2j+1}e^{-s^2}}{\Gamma(1 + j, s^2)}, 
	\\
p_{\textsc{nnn}}^{\rm (A)}(s) &=&E^{\rm (A)}_{\textsc{nn}}(s)      \sum_{j,k=1; k \neq j}^{\infty}  
\frac{\gamma(1+j, s^2)}{\Gamma(1 + j, s^2)}
\frac{2s^{2k+1} e^{-s^2}}{\Gamma(1 + k, s^2)}	,
\nonumber\\
	\label{ANNN}
\end{eqnarray}
where 
\begin{eqnarray}
\gamma(1 + k, s^2)&=&\int^{s^2}_0 t^k e^{-t}d t\ ,\\
\Gamma(1 + k, s^2)&=&\Gamma(1+k)-\gamma(1 + k, s^2)\ ,
\end{eqnarray}
are  the lower  and upper incomplete
Gamma function. Furthermore,
\begin{equation}
\label{gap}
E_{\textsc{nn}}^{\rm (A)}(s)= 	\prod_{j=1}^{\infty}\frac{\Gamma(1 + j, s^2)}{\Gamma(j+1)}
\end{equation}
is the so-called gap probability in class A to find an 
eigenvalue at the origin and the closest non-zero complex eigenvalue at radial distance $s$.
For finite $N$ the sums and product in \eqref{ANN}-\eqref{gap} extend only to $N-1$.  The product converges very rapidly and can be truncated at, say $N=20$. 
It is also clear form this formulas that $N=2$ is not a good approximation here, as was already noted in \cite{grobe1988quantum}.
Both spacing distributions \eqref{ANN} and \eqref{ANNN} are normalised to unity. However, the first moment still has to be rescaled to unity, by numerically computing 
$s_1=\int_0^\infty s \,	p_{\textsc{nn}}^{\rm (A)}(s) ds$, and then rescaling  	$p_{\textsc{nn}}^{\rm (A)}(s) \to 	s_1 p_{\textsc{nn}}^{\rm (A)}(s_1s)$. After doing the same rescaling of \eqref{ANNN} with the same factor $s_1$, the resulting distributions are shown in Fig. \ref{Fig:3RMTs} left and right, middle curves, 
see also Fig. \ref{Fig:2d-statistics}(b) left, respectively right, class A being given by the curve with the intermediate maximum.

\subsection*{Class AII$^\dag$ - definition and fits to NN and NNN}
\begin{figure}[]
	\includegraphics{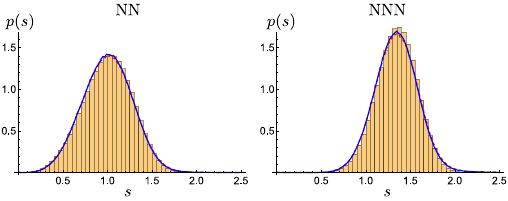}	
	\caption{Comparison of the NN (left) and NNN (right) spacing distribution for $500$ matrices of size $2N=5000$ of the symmetry class AII$^\dagger$ (histogram), with the best fit of the 2dCG data for $\beta=2.5$ (red dashed curve) and $\beta=2.6$ (blue curve).
The comparison to NNN is new here.  	
	}
	\label{Fig:AII+}
\end{figure}

We turn to the definition of the symmetry class AII$^\dagger$. The ensemble consists of non-Hermitian matrices of complex self-dual  matrices $J$ with quaternion elements, that we represent by $2N\times 2N$ complex matrices. They satisfy 
\begin{equation}
    J = \Sigma_N J^T \Sigma_N,\quad P_N^{(\rm{AII}^\dag)}(J)=C_N e^{-\textrm{Tr}(JJ^\dagger)}, 
\end{equation}
with Gaussian distribution of elements, normalisation $C_N^{-1}=(\pi/2)^{2N^2-N}$, and 
$\Sigma_N$ the skew-Hermitian metric 
\begin{equation}
    \Sigma_N = \begin{pmatrix}
        0 & -i \mathbb{1}_{N\times N} \\
        i \mathbb{1}_{N\times N} & 0 
    \end{pmatrix}.
\end{equation}
The NN and NNN spacing distributions have to be determined numerically again and are shown in Fig. \ref{Fig:AII+} as histograms.

\begin{table}[h]\centering
	\caption[]{List of standard deviations $\sigma$ and Kolmogorov-Smirnov distances $d$ (both in units $ 10^{-2}$) 
		between NN and NNN spacings of AII$^\dagger$ and 2dCG,  
		with $\beta\in[2.3,2.8]$ in steps $0.1$. 
		For NN the best fit is with $\beta=2.6$ for both distances, as is for the NNN for $\sigma$, whereas $d$ favours $\beta=2.5$ for NNN (bold numbers). 
		}\vspace{2mm}
	\begin{tabular}{|c|c|c|c|c|c|c|c|}
		\hline
		fitted &$\beta$  & 2.3 & 2.4 & 2.5 & 2.6 & 2.7& 2.8\\
		\hline
  \hline
		\multirow{2}{*}{NN}&	$\sigma$ & 3.4 & 2.8 & 2.4 & \bf{2.3} & 2.4 & 3.0 \\
		\cline{2-8}
		&$d$ & 1.4 & 1.1 & 1.0 & \bf{0.9} & 1.0 & 1.2 \\
		\hline
		\hline
		\multirow{2}{*}{NNN}&$\sigma $& 5.5 & 4.6 & 4.1 & \bf{3.9} & 4.3 & 4.9 \\
		\cline{2-8}
		&$d$ & 2.8 & 2.1 & \bf{1.6} & 1.9 & 2.4 & 2.8\\
		\hline
	\end{tabular}
	\label{Tab:NN and NNN AII dag 2dC}
\end{table}

Also shown is a fit of the two respective spacing distributions to $\beta$ in the 2dCG. The corresponding Kolmogorov-Smirnov distances $(d)$ and standard deviations $(\sigma)$ are given in Table \ref{Tab:NN and NNN AII dag 2dC}, with best fits marked in bold. For the NN the value $\beta=2.6$ is slightly favoured, which was also reported in \cite{akemann2022spacing}. For NNN, 
 $\beta=2.6$ again gives the best fit according to the standard deviations, whereas 
$\beta=2.5$ is best for the Kolmogorov-Smirnov distant. This shows that also the second next eigenvalue of the 2dCG with its locally logarithmic repulsion gives a consistent approximation of this random matrix symmetry class AII$^\dag$. 

For comparison with the quality of fits in class A, where we have analytical predictions, we independently  ran numerical simulations with matrices of the same size, see Table \ref{Tab:NN and NNN A 2dC} at the end of Subsection \ref{Sec:numerical}.

\vfill


\bibliography{references}
\end{document}